\documentclass[prb,preprint,amsmath,amssymb]{revtex4}
\usepackage{color}
\begin{document}
\author{Kevin Leung,$^*$ Laura C.~Merrill, and Katharine L.~Harrison}
\affiliation{Sandia National Laboratories, MS 0750, Albuquerque, NM 87185\\
$^*${\tt kleung@sandia.gov}}
\date{\today}
\title{Galvanic Corrosion and Electric Field in Lithium Anode
Passivation Films: Effects on Self-Discharge}

\input epsf.sty
%\ssp
 
\begin{abstract}

Battery interfaces help govern rate capability, safety/stability, cycle
life, and self-discharge, but significant gaps remain in our understanding
at atomic length scales that can be exploited to improve interfacial
properties.  In particular, Li partially plated on copper current collectors,
relevant to the anodeless, lithium metal cell which is a holy grail of
high density energy battery research, has recently been reported to undergo
galvanic corrosion and exhibit short shelf lives. We apply large scale Density
Functional Theory (DFT) calculations and X-ray photoelectron spectroscopy
to examine the reaction between the electrolyte and Li$|$Cu junctions coated
with uniform, thin electrolyte interphase (SEI) passivating films at two applied
voltages.  The DFT galvanic corrosion calculations show that electrolyte
degradation preferentially occurs on Li-plated regions and should lead to 
thicker SEI films.  We find similarities but also fundamental differences
between traditional metal localized pitting and Li-corrosion mechanisms due
to overpotential and ionic diffusion rate disparities in the two cases.
Furthermore, using the recently proposed, highly
reactive lithium hydride (LiH) component SEI as example, we distinguish
between electrochemical and chemical degradation pathways which are 
partially responsible for self-discharge, with the chemical pathway found
to exhibit slow kinetics.  We also predict that electric
fields should in general exist across natural SEI components like LiH,
and across artificial SEI films like LiI and LiAlO$_2$ often applied to
improve battery cycling.  Underlying and unifying these predictions
is a framework of DFT voltage/overpotential definitions which we have derived
from electrochemistry disciplines like structural metal corrosion studies; our
analysis can only be made using the correct electronic voltage definitions.

\end{abstract}
 
\maketitle
 
\section{Introduction}
\label{introduction}

Lithium (Li) metal anodes represent one of the holy grails of battery research
because of its high gravimetric capacity advantage.\cite{intro}  So far, high
coulombic efficiency\cite{cycling} remains an elusive goal, especially in
liquid electrolyte-based batteries which are our focus.  Recently
self-discharge,\cite{selfdischarge,mengsd,cuisd,galvanic1,galvanic2,merrill,galvanic3}
whereby the stored energy in a battery dissipates without generating a current
in the external circuit, has been identified as another potential issue,
especially in anodes with three-dimensional architectures.\cite{galvanic2}  
Fast self-discharge has been attributed to ``galvanic corrosion'' which occurs
in lithium (Li) plated on copper (Cu) current collectors.  Vastly different
self-discharge rates have been reported,\cite{cuisd,galvanic1,galvanic2,merrill}
depending on the amount of Li plated; the degree of self-discharge
reversibility also varies.  This is surprising because Cu current collectors
are widely used, and are usually in contact with Li on one side and the
electrolyte on the other side, yet galvanic corrosion has seldom been noted.

The lithium galvanic corrosion phenomenon offers a unique opportunity for
comparison between traditional (e.g., localized pitting) metal
corrosion\cite{corrbook} and degradation in lithium battery
electrodes.\cite{pccp,kostecki}
Metals like aluminum (Al) have low reduction potentials that lie outside the
electrolyte (e.g., water) stability window, and has to be passivated with
native oxide films.  Galvanic processes occur when a more
electropositive metal is in contact, leading to $e^-$ transfer from and
enhanced dissolution of the less electronegative metal.\cite{jced} These
aspects have profound similarities with Li metal and other Li-battery
anodes\cite{pccp} which are metastable against most electrolytes and are always
passivated by solid electrolyte interphase (SEI)
films.\cite{xu2004,qi_review,vegge,canepa}
Even if Li is scraped clean in a glove box, it is ``dirty'' in that it
immediately getters any oxygen or water in the atmosphere and forms a
oxide coating to passivate itself (supporting information 
document, S.I., Sec.~S1).  In many cases the SEI film is generated by
electrochemical reduction of the electrolyte during plating, and/or by
subsequent chemical reactions of the initial SEI or electrolyte with the plated
lithium metal.\cite{xu2004,sei,qi_review} In other cases, artificial SEIs
(aSEIs) are coated directly on Li metal, or first on the current collector
prior to Li plating,\cite{kozen,elam} to improve cycling and self-discharge
behavior via improvement of SEI passivation properties or anode morphology.
Artificial passivation is also used to protect structural metals from
corrosion.\cite{corrbook}  Other links between traditional corrosion research
and battery studies include stress corrosion cracking\cite{sjharris} and
current collector degradation.\cite{curtiss}  

There are also fundamental differences between batteries and Al/steel-corrosion,
so much so that very notions of ``galvanic corrosion,'' and even ``corrosion,''
in a Li anode context need to be re-examined.  For example, Al and steel
pitting onset occurs when the passivating oxide films break down,
partly because of the slow metal cation diffusion rates in the oxide.  Hence
the anodic Al oxidation step is corrosion-limiting, and this is accelerated
at higher voltages.  \color{black} Modeling of corrosion of structural corrosion
is challenging because corrosion usually involves
overpotentials against metal plating (e.g., the mininum Al pitting
potential is $\sim$ -0.5~V vs.~SHE,\cite{pittingvoltage} much higher than
the -1.66~V~vs.~SHE of Al$^{3+}$ + 3 $e^-$ $\rightarrow$ Al(s)), spatial
segregation of cathodic and anodic processes, and kinetics (not only
thermodynamic) considerations.\cite{corrbook} \color{black}  
Li$^+$ diffusion is sufficiently fast in its SEI that overpotentials are less
relevant, and that cathodic reactions (electrolyte reduction), accelerated at
low voltages, might be limiting; we will use DFT and measurements to demonstrate
this point.  Despite these caveats, we propose that synergistic studies of
metal corrosion mitigation and lithium anode protection will yield
significant, cross-cutting benefits to both fields.

\begin{figure}
\centerline{\hbox{ \epsfxsize=5.50in \epsfbox{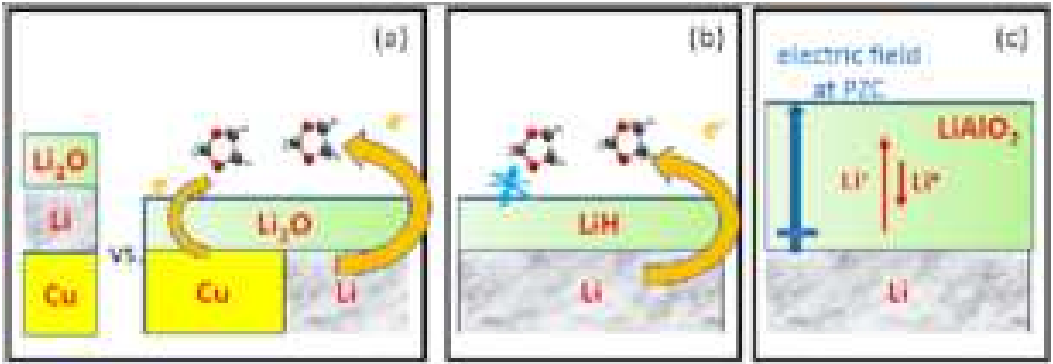} }}
\caption[]
{\label{fig1} \noindent
Schematics illustrating
(a) galvanic corrosion, with solvent (DOL) electrochemical composition 
    favored on plated Li metal regions over Cu regions at same SEI
    thickness;
(b) solvent chemical decomposition on LiH surfaces vs.~electrochemical
    decomposition on LiH-coated Li surfaces;
(c) predicted electric field at LiAlO$_3$ interface near
    potential-of-zero-charge (PZC) conditions leads to prediction of
    asymmetric Li$^+$ plating/stripping rates.
}
\end{figure}

We hypothesize that galvanic corrosion-like effects in Li anodes occur due to
quasi-three-point phase boundaries between the Cu current collector, the plated
Li, and the electrolyte -- with the electrolyte separated from the two metals
by a nascent SEI.  In this work, we use DFT slab models with up to 3000~atoms
(sizes almost unprecedented in previous battery studies) we investigate Cu$|$Li
metal junctions at the nanoscale.  50~\% and 100\% coverages of Cu-current
collector by lithium metal are considered (Fig.~\ref{fig1}a).  Here we 
exploit a previous established framework in the use of Density Functional
Theory (DFT) to model battery interfaces.\cite{pccp} These advances are based
on electrochemical modeling paradigms often developed outside of battery
research;\cite{pccp,otani,peterson,iceland,mira,dabo1,marzari,filhol,pham} have
been applied to model Al-corrosion;\cite{corrosion} include electrochemical
voltage/overpotential definitions; and reveal complexities like electric
field\cite{maier} and contact potential\cite{pccp,gb,corrosion,marks} effects
seldom addressed in the battery modeling literature.  They go far beyond
thermodynamics/phase diagram analysis of interfacial stability, although the
latter approach is important too.\cite{wolve,siegel,vdv,ceder1,islam1,phase}
By computing work functions and modeling electrochemical reduction reactions
of solvent molecules, we demonstrate the critical role played by spatial
inhomogeneity.  Solvent decomposition consumes Li metal atoms and ultimately
turn them into Li$^+$.  This consumes Li inventory and leads to irreversible
losses and self-discharge. 

We focus on the fundamental chemistry and physics of solid-solid interfaces
associated with thin-film coated (``dirty'') battery electrodes.  The nascent,
uniformly-thick SEI films in our models are thin enough that they are not
considered final structures, but can still grow and evolve.  This allows us to
concentrate on intermediate stage SEI growth and
evolution\cite{evolve1,evolve2,evolve3,evolve4} on top of native 
or engineered\cite{gallant1} oxides generally present on Li metal anodes,
unlike prevalent modeling work on pristine Li(s)/liquid electrolyte interfaces
mostly relevant to the initial stages of SEI
growth.\cite{schmickler,qi_review,note0,highconc,highconc2,wise,nasa,baskin,lii}
\color{black}
We also refrain from making {\it a priori} assumptions about our model systems
reaching equilibrium between electronic and ionic properties, unlike
previous work which do not address overpotentials.\cite{gerischer,swift}
DFT time scales are short compared to experiments (Supporting Information,
SI Sec.~S6) which can lead to unintentional overpotentials.\cite{pccp}  We
match possible or experimentally-determined overpotential values during
self-discharge/corrosion to DFT (electronic) voltages by tuning the interfacial
structure.  In other words, we focus on voltage-function relations, not
structure-function relations which are less pertinent because
atomic length-scale experimental interfacial structures have seldom been
elucidated.  Kinetics considerations are addressed using both explicit reaction
barrier calculations, and separation-of-time-scale approximations. \color{black}

Understanding the chemical and electrochemical stability of SEI film components
and their reactions with the electrolyte is also relevant to predicting Li
``corrosion'' processes.  Our DFT framework and electronic voltage
definitions\cite{pccp} also permit us to distinguish electrochemical
vs.~chemical electrolyte degradation reactions, both of which can contribute
to self-discharge.  (Here ``chemical'' refers to reactions that can occur
without an electron source, and includes both energetic and kinetic
considerations.) To this end, we adopt as a test case the reactive
lithium hydride (LiH), which has surprisingly been reported to exist in the
SEI.\cite{lih1,lih2,lih3}  We hypothesize that LiH, which is thermodynamically
unstable against water, oxygen, and most organic compounds, must exhibit
kinetically-limited reactions in order to persist and be observed
in the SEI (Fig.~\ref{fig1}b).  

Our work on LiH reveals other complexities that can potentially be exploited
to reduce self-discharge.  The contact potential between LiH and Li metal is
significantly lower than that between Li metal and well-known SEI components
LiF or Li$_2$O.\cite{pccp}  This leads us to survey other materials like LiI
and LiAlO$_2$ which have been proposed as aSEI
materials,\cite{lii,jacs,elam,kozen} and which reveal an even wider range of
contact potential values.  Since the contact potential plus mathetmatical
integration of the electric field across the SEI and electrolyte add up to
the applied potential,\cite{pccp} this implies that different field strengths
must exist inside different SEI/aSEI films at a given voltage.  The fact
that battery electrode interfaces have surface charges and exert electric
fields should be expected because many electrified interfaces only exhibit
one ``potential of zero charge'' (PZC) and should have net charges away from
the PZC; neverthless electric fields in the SEI have mostly been neglected
until recently.\cite{maier,filhol,pccp}  We illustrate this using cross-film
Li$^+$ transport in LiAlO$_2$ coated on Li metal (Fig.~\ref{fig1}c), and
propose that interfacial electric field engineering will have beneficial
properties.\cite{yada,ye2021} Our models represent no-current snapshots of SEI
growth\cite{maier} and/or Li self-discharge,\cite{galvanic1,galvanic2,merrill}
not the entire dynamic ``corrosion'' process.  Nevertheless, they illustrate
useful new concepts that should motivate future experiment work. 

\section{Computational Methods}

%\vspace*{0.1in}
\subsection{\it {\bf DFT Details}} 
%\noindent {\it {\bf DFT Details}}  \\

All DFT calculations are conducted under T=0~K ultrahigh vacuum (UHV)
conditions, using periodically replicated simulation cells and the Vienna
Atomic Simulation Package (VASP) version 5.3.\cite{vasp1,vasp1a,vasp2,vasp3}
A 400~eV planewave energy cutoff and a 10$^{-4}$~eV convergence criterion
are enforced.  Most calculations apply the Perdew-Burke-Ernzerhof
(PBE) functional.\cite{pbe}  In the S.I., HSE06 is used for
spot checks.\cite{hse06a,hse06b,hse06c}   The standard dipole correction is
applied.\cite{dipole}  It eliminates spurious coupling between periodic
images in the $z$-direction.  However, no standard correction currently exists
for dipole/dipole-image interactions in the $x$-$y$ plane; the consequences
will be discussed below.  Transition state energies or barriers ${\Delta E^*}$
are computed using the climbing image nudged elastic band method.\cite{neb}

The two- and multilayer SEI models\cite{aurbach_two_layer,li_dual_layer,shinoda}
suggest that inorganic products coat the active anode material, separating
it from the thicker, amorphous, and porous organic/polymeric layer outside.
More recently, cryogenic transmission electron microcopy (cryo-TEM) images
suggest an amorphous matrix in contact with Li metal, in which Li$_2$O and LiF
nanoparticles are embedded in some electrolytes.\cite{meng,cui,jungjohann}
While the chemical
identity of the armophous phase has not been revealed by TEM, candidates that
are stable against Li metal are extremely limited.\cite{batt}  A recent
demonstration of reactions between Li$_2$CO$_3$ and Li metal surfaces suggests
this matrix may contain graphitic carbon with Li intercalated arising from
multistep electrochemical reduction of organic electrolyte and their
intermediate products, including Li$_2$CO$_3$.\cite{umd}  Since graphitic
carbon is metallic and does not block $e^-$, here we focus on
Li$_2$O\cite{gallant1,shinoda} and other insulating inorganic SEI products as
the inner SEI layer (S.I. Sec.~S1).  \color{black} The organic SEI outside the
inorganic SEI has complex, heterogeneous compositions and structure, and likely
exhibits a low dielectric constant.  As before,\cite{pccp,gb} we approximate
the outer organic SEI as a vacuum.  Our implicit assumption is that, to first
order, all the electric fields and voltage drops are contained within the
inorganic SEI (or aSEI).  To the extent that the low dielectric constant
organic SEI is common to all interfacial systems considered in this work, its
effects on DOL reactivity on different Li$_2$O-covered surfaces should largely
cancel.\color{black}
Organic SEI and liquid electrolyte effects will be investigated in future
work.\cite{otani} Due to the approximations used in this work (thickness of
material slabs, finite DFT accuracy, omission of organic SEI and electrolyte
regions), the properties predicted are semi-quantitative.  However, the
well-defined approximations in our models provide a path towards
future systematic improvements.

Regarding bulk inorganic SEI/aSEI component crystal structures, we obtain a
4.013~\AA\, lattice constant in the 4 formula unit, cubic LiH unit cell.
The (001) and (110) facets are found to exhibit surface energies of
0.33 and 0.74~J/m$^2$, respectively.  Henceforth we focus on (001).  LiI
has a similar cubic structure, and a 6.025~\AA\, cubic unit with 4 formula
units.  The 4-formula unit LiAlO$_2$ unit cell has 
5.30$\times$6.35$\times$4.96~\AA$^3$ dimensions.\cite{jacs}

The lowest energy Li surface, Li(001), is adopted.  The initial registries at
the interfaces are selected as follows.  (Here ``registry'' refers to how the
surface unit cells of the two materials are aligned or spatially offset in the
$x$-$y$ plane.) LiI, Li$_2$O, and LiH exhibit weak interactions with Li metal
surfaces (although the contact potential can still be large); the precise
registry is not critical, and only one choice each is
used.  Unless otherwise noted, one Li atom is inserted directly below
each anion on the inner surface of the inorganic films (i.e., I$^-$ in
LiI, O$^{2-}$ in Li$_2$O, and H$^-$ in LiH).  This Li ``interlayer''
has been found to lower the electronic voltage (${\cal V}_e$, Eq.~\ref{eq2}
below) and is either thermoneutral or exothermic at some interfaces, after
subtracting the Li metal cohesive energy (i.e., ${\cal V}_i$$>$0~V, 
Eq.~\ref{eq1} below).\cite{gb} Other details of the interface models are given
in Table~S3 (S.I.).  When doubling or quadrupling simulation cells for system
size convergence purposes, we start from the converged configurations of
the baseline cells (S.I. Table~S3).

For LiAlO$_2$, where a significant binding energy with Li metal is expected,
we have attempted a different procedure.  We displace the Li metal slab
against the oxide in a $x$-$y$ grid of increment 0.5~\AA, optimize the
configurations, and select the most stable of the 81 resulting registries.  The
computed standard deviation in interfacial energy is $\pm$0.02~J/m$^2$.  After
this grid search, it is necessary to further conduct {\it ab initio} molecular
dynamics (AIMD) T=450~K for 9~ps, followed by a quench from T=450~K to
T=50~K over 4~ps and then an energy minimzation calculation, to obtain the
final Li/LiAlO$_2$ configuration.  Without AIMD annealing, the introduction
of Li-vacancies, needed to model Li diffusion energies, leads to very
significant reconstructions at the interface.  Experimentally, aSEI/Li(s)
interfaces are in many cases created by first forming the aSEI on Cu and then
plating Li metal underneath the aSEI.  How much of this room temperature
process is qualitatively consistent with our ``cold press,'' ``light annealing''
short AIMD trajectory approach will be considered in future work.  

The Li$|$Cu junction is first optimized as a 
$x$$\times$10.16$\times$36.00~\AA\, simulation cell with a Cu$_{352}$Li$_{195}$
stoichiometry and two metal-vacuum interfaces in the $z$ direction.
The optimal $x$ is found to be 34.17~\AA.  Then the Li$_2$O (111) oxide
is place on top of it while straining the oxide and metal slabs in the
$x$-$y$ plane to fit their average surface supercell dimensions.  
The optimization technique used herein locates local minima, not the globally
most favorable configuration.  We do not observe Li/Cu alloying.

%\vspace*{0.1in}
\subsection{\it {\bf Voltages and Electric Fields}} 
%\noindent {\it {\bf Voltages and Electric Fields}} \\

Li anodes emit and consume both ions and electrons, so two voltage definitions
can be introduced.  We define the ionic voltage, ${\cal V}_i$
as\cite{voltage_prb}
\begin{equation}
{\cal V}_i=[(E_{n_{\rm M}}-E_{n'_{\rm M}})/(n_{\rm M}-n'_{\rm M})
	- \mu_{\rm M}]/|e|,
							\label{eq1}
\end{equation}
where $E_{n_{\rm M}}$ is the total energy of the simulation cell with an
electrode with $n_{\rm M}$ M atoms, $\mu_{\rm M}$ is the M chemical
potential in its bulk metal phase, and $|e|$ is the electronic charge.
${\cal V}_i$ is the energy of inserting an Li atom at the most favorable
Li-insertion site, referenced to Li metal cohesion energy, divided by $|e|$.
The electronic voltage ${\cal V}_e$ is
\begin{equation}
{\cal V}_e = \Phi/|e| -1.37~V, \label{eq2}
\end{equation}
where $\Phi$=$(E_{\rm vacuum}$-$E_{\rm F}$) is the work function,
$E_{\rm vacuum}$ is the vacuum levels in the same simulation cell in which
the Fermi level $E_{\rm F}$ is computed, and 1.37~V is related to the Trassati
relation,\cite{trasatti} adapted to the Li$^+$/Li(s) reference.  Physically,
${\cal V}_e$ is to be compared with the voltage measured or controlled using
potentiostats, with necessary caveats about the different time scales involved
in DFT and experimental overpotentials given in the S.I. (Sec.~S6).
${\cal V}_e$ depends on the $k$-point sampling and material slab thicknesses,
and reported values herein are precise to $\sim$0.1~V.  We stress that
${\cal V}_i$ and ${\cal V}_e$ values are referenced to Li$^+$/Li(s). 

${\cal V}_e$ and $\Phi$ are altered by the effective dipole moment in the
surface film and at material interfaces via
\begin{equation}
\Delta \Phi = 4 \pi \sigma d . \label{eq3}
\end{equation}
Here $\sigma$ is the surface density of a uniform point dipole sheet, $d$ is
the average dipole magnitude, and atomic units are used.  We define
${\cal E}_e$ as the electric field across the insulating film, not including
the contact potential.

\color{black}
``$i$-equilibrium'' is said to be established when ${\cal V}_e$=${\cal V}_i$.
When ${\cal V}_e$$\neq$${\cal V}_i$, ${\cal V}_e$ is offset by the
``overpotential'' value from ${\cal V}_i$ and is away from $i$-equilibrium.
In this sense, ${\cal V}_e$ is not an independent ``prediction''; it should
be treated as a constraint, pegged to experimentally determined potentials
or overpotentials.  Overpotentials are generally determined by processes
too slow for DFT to predict. \color{black}

We also define a local electrostatic potential (or local vacuum level)
${\phi}(x)$, obtained by integrating over the 3-dimensional electrostatic
potential $\phi (x,y,z)$
\begin{equation}
{\phi}(x) = 1/[L_y (z_2-z_1)] \int_{z1}^{z2} dz \int_0^{L_y} dy \phi(x,y,z).
\end{equation}
Here $L_y$ is the cell dimension in the $y$ direction, 
\color{black}
and $z_1$ and $z_2$
bracket the vacuum region in the slab between 3.0~\AA\, and 6.0~\AA\, above
the mean position of the top atomic layer of the slabs.  This is a qualitative
guide to the tendency of electron transfer to a range of $z$ values roughly
commensurate with the size of an adsorbed atom.  $\phi(x)$ is not physical
or measurable.  Only relative values of ${\phi}(x)$ are relevant. \color{black}

Within ground state DFT, each metal, metal alloy, and metal junction slab has a
single $E_{\rm F}$, and hence a unique ${\cal V}_e$, even if local variations
in ${\phi}(x)$ can occur.\cite{gb} In DFT slabs, there can be two ${\cal V}_e$
per metal slab because the front and back sides (coated with surface films or
not) can have different vacuum levels which are split by the artifial dipole
sheet that keeps the vacuum free of electric fields.\cite{dipole} We only
report ${\cal V}_e$ on the side of the coated Li slab, unless otherwise stated.

We consider models in which nascent passive films of thickness $<$15~\AA\,
exist on metal surfaces.  This thickness should still allow electron
leakage through the film.\cite{gb2}  Hence we do not consider Marcus theory
for long-range $e^-$ transfer,\cite{jacs,gb2,marcus1,marcus2,miller} 
or non-adiabatic effects.\cite{marcus3}  Finally, there are fundamental
disconnects between the battery and other electrohemical computational
literature concerning voltage definitions and electric fields/electric double
layers (EDL), at the interfaces of both liquid and solid electrolytes.\cite{corrosion,pccp,maier,voltage_prb,borodin,latz,tateyama,tateyama2,janek,macdonald81,field1,field2,gewirth,aizawa,luntz,liang2018,qi}
The S.I. (Sec.~S5) attempts to reconcile different viewpoints.

\section{Results and Discussions}

%\vspace*{0.1in}
%\noindent {\it {\bf DOL Reductive Decomposition on Coated Li Metal}}
\subsection{\it {\bf DOL Reductive Decomposition on Coated Li Metal}}

\begin{figure}
\centerline{\hbox{ (a) \epsfxsize=2.00in \epsfbox{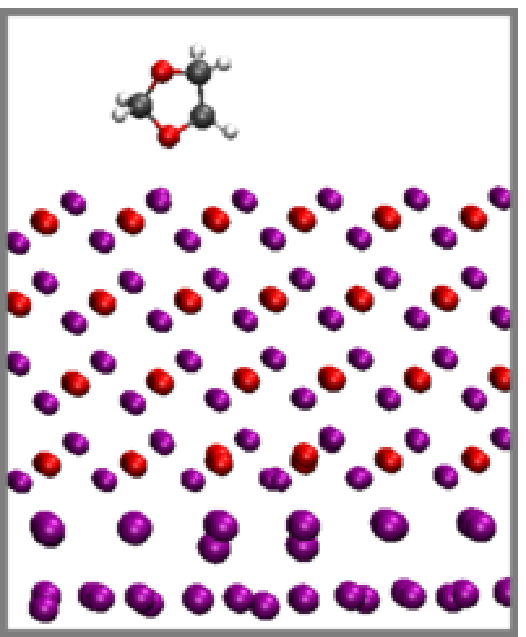} 
		   \epsfxsize=2.00in \epsfbox{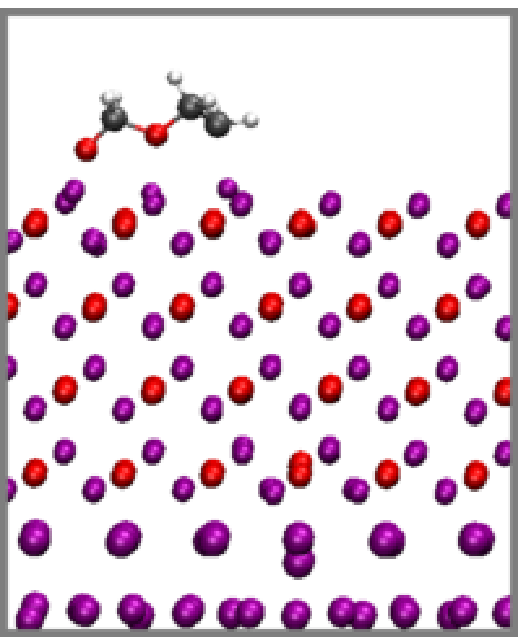} (b) }}
\centerline{\hbox{ (c) \epsfxsize=2.00in \epsfbox{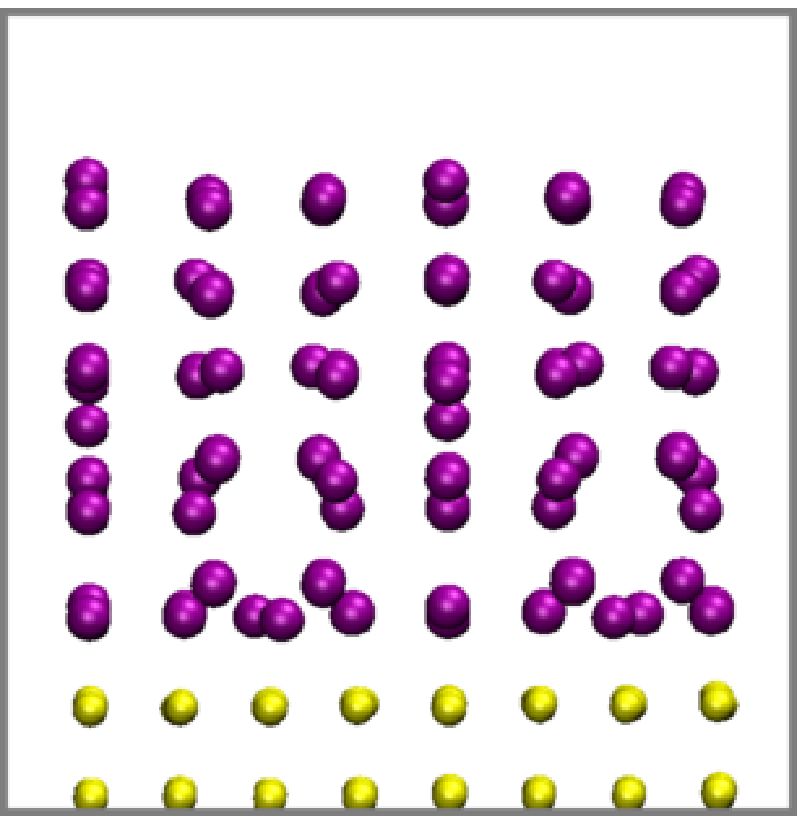} 
		   \epsfxsize=2.00in \epsfbox{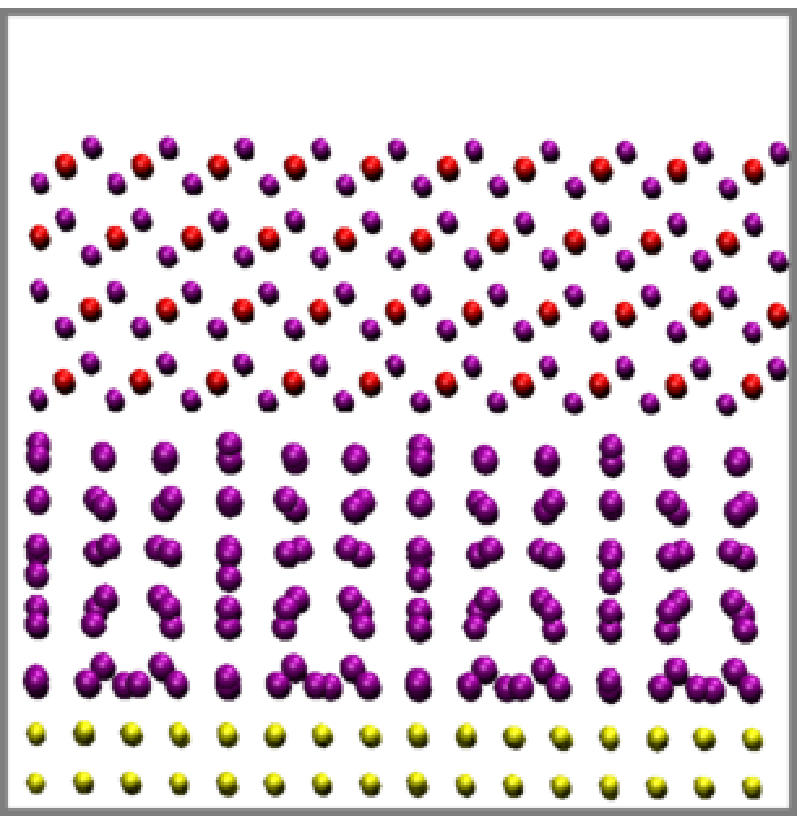} (d) }}
\caption[]
{\label{fig2} \noindent
(a)-(b) Intact and decomposed DOL on Li$_2$O (111) surface coating
Li metal at the bottom.  (c) Li(001)/Cu(001) interface in a vertical
stack; (d) Li$_2$O(111)/Li(001)/Cu(001) interfaces.  Cu, Li, O, C, and H
atoms are depicted as yellow, purple, red, grey, and white spheres.  
${\cal V}_e$=0.05~V in both Fig.~\ref{fig2}a~and~\ref{fig2}d.
}
\end{figure}

To demonstrate the basic science and DFT system size effects involved, we
first consider the reaction of 1,3-dioxolane (DOL) on pure Li$_2$O-coated Li
metal surfaces.  DOL is a cyclic ether frequently used as a co-solvent
in Li metal anode batteries.\cite{dol1}  It is chosen as the sole adsorbed
solvent for computational convenience; its ring structure requires less
exploration of the configuration space than linear molecules like
1,2-dimethoxyethane (DME).  In the S.I. (Sec.~S4), we estimate that the DOL
reduction potential in the electrolyte is below 0.0~V vs.~Li$^+$/Li(s). This
suggests that DOL electrochemical reduction most likely initiates on the
electrode surface, where $e^-$ transfer to DOL should be more favored via
surface stabilization effects.  
Note that the electrolyte anion is in fact
more readily reduced than the ether solvent, although organic SEI components
are also found (S.I. Sec.~S1).  The higher anion reactivity makes it
more difficult to distinguish ${\cal V}_e$ effects in our proof-of-principle
calculations, and DOL is chosen instead.  \color{black} In the following we focus
on concerted electron transfer and DOL decomposition reaction on inorganic
SEI surfaces; the DOL is assumed to be embedded in the low-dielectric
organic SEI, not the solvent. We do not focus on its reduction potential,
only the initial DOL bonding-breaking event.  The chemical constituent of
DOL-based SEI has not yet been completely elucidated.\cite{kang_dft,bal_dol}
Thus, DOL is meant as a stand-in for the organic SEI in addition to
representing the solvent.  \color{black} 

In Fig.~\ref{fig2}, a $\sim$10~\AA\, thick Li$_2$O (111) layer covers the Li
(001) surface,\cite{gb} giving ${\cal V}_e$$\sim$0.05~V vs.~Li$^+$/Li(s)
(Table~\ref{table1}),\cite{gb,pccp} which is very close to the Li plating
potential.  Fig.~\ref{fig2}a-b depict an intact and a decomposed DOL on a
$\sim$10~\AA\, thick Li$_2$O film on the Li (001) metal surface.  Bader charge
analysis\cite{bader} reveals that the DOL with one broken bond is doubly
reduced.  Table~\ref{table1} shows that, in the largest simulation cell
considered, the energy released is $\Delta E$=-1.72~eV, while the reaction
barrier is $\Delta E^*$=+0.67~eV, in reasonable agreement with the $\sim$0.6~eV
computed in liquid using a different DFT method.\cite{kang_dft}  The latter
translates into a sub-second reaction time scale using standard kinetic
equations with a 10$^{12}$/s prefactor.\cite{pccp} As battery interfaces are
usually governed by kinetics, not thermodynmics, $\Delta E^*$ is more
significant than $\Delta E$ as long as the converged exhibits $\Delta E$$<$0 
(exothermic reaction).  

Next we discuss the convergence of $\Delta E$ and $\Delta E^*$ with system
size.  Adsorption of an intact DOL does not entail $e^-$ transfer and should
not change ${\cal V}_e$ significantly.  For the decomposed DOL, the two $e^-$
transfer from Li metal to the molecule leads to an artificial increase of the
dipole moment in the charge-neutral simulation cell, and a significant increase
in ${\cal V}_e$ (Eq.~\ref{eq2}-\ref{eq3}).  Thus $\Delta {\cal V}_e$, defined
as the difference in ${\cal V}_e$ between the slabs with decomposed and intact
DOL, is a measure of system size dependence; it should converge to zero in
the limit of infinite system size.  $\Delta {\cal V}_e$ can be as large as
2.08~V in slabs with small lateral dimenions (Table~\ref{table1}).  As the cell
size increases, ${\cal V}_e$ drops to 0.14~V. Simultaneously, $\Delta E$
increases in magnitude from -0.85~eV in a 1$\times$1 cell
(17.0$\times$9.9~\AA$^2$ lateral area, S.I. Table~S3) to -1.72~eV in a
2$\times$2 cell (34.1$\times$19.7~\AA$^2$, 1427~atoms).  In the largest cell,
$\Delta E$ has apparently converged to within $\sim$0.17~eV.  $\Delta E^*$ is
more costly to compute than $\Delta E$.  Fortunately, the transition state
associated involves the transfer of $<$2~$e^-$ from the metal to the DOL, and
the convergence of $\Delta E^*$ with respect to system size is faster
than for $\Delta E$ (Table~\ref{table1}).

We also consider this DOL reaction in the absence of the Li metal
underneath the Li$_2$O layer by removing the Li slab and re-optimizing.  The
atomic configurations look similar to those in Fig.~\ref{fig2}a-b and are not
shown.  $\Delta E$ jumps to +2.43~eV.  This suggests that, without an $e^-$
source, DOL decomposition reaction does not occur, unless other processes,
such as cation-induced polymerization of DOL, occur.  Another configuration
where the dangling CH$_2$ group reacts with a surface O-atom on Li$_2$O slabs
without a Li metal slab underneath, and thus do not involve $e^-$ transfer
from Li metal, is energetically but not kinetically favorable,
and is not shown in Fig.~\ref{fig2}.  See the SI Sec.~S9 for details.
As another point of reference, on pristine Li (001) surfaces, DOL
readily breaks one C-O bond to release $\Delta E$=-3.67~eV with a barrier
of $\Delta E^*$=0.74~eV (S.I. Sec.~S4); even more energy is released during
subsequent DOL bond-breaking events.  However, as we have stressed, pristine Li
metal is too reactive to exist in batteries (S.I. Sec.~S1).  These calculations
do not involve $e^-$ transfer across an insulating oxide film and should not
incur large system size effects.\cite{pccp}

\begin{figure}
\centerline{\hbox{ \epsfxsize=3.60in \epsfbox{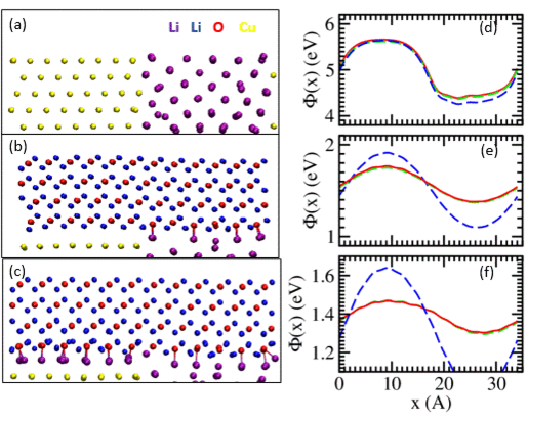} }}
\centerline{\hbox{ \epsfxsize=1.80in \epsfbox{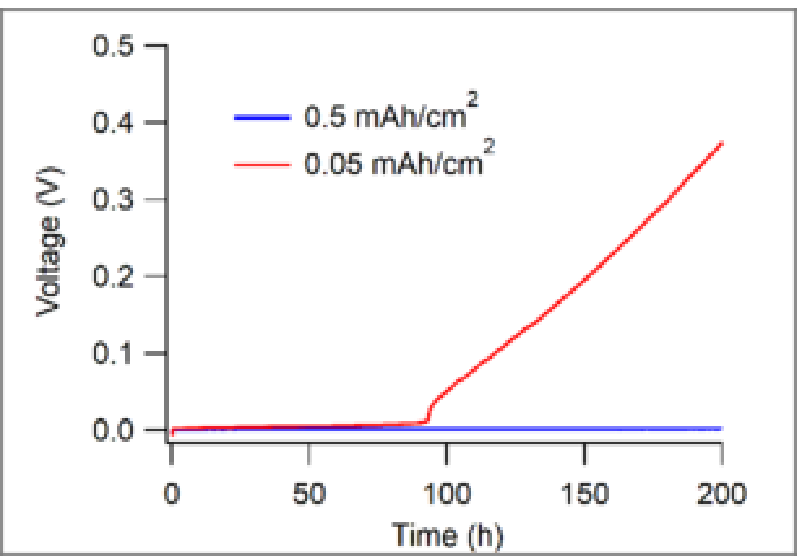} 
		   \epsfxsize=1.85in \epsfbox{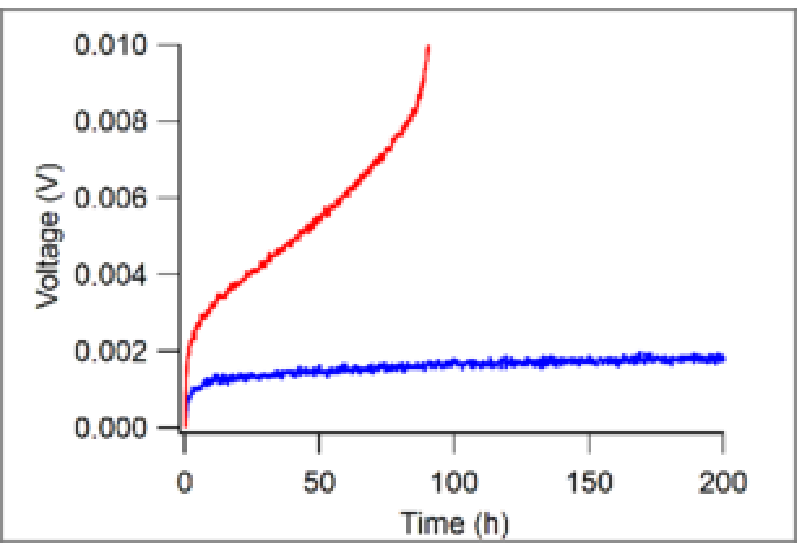} }}
\centerline{\hbox{  (g) \hspace*{1.80in} (h) }}
\caption[]
{\label{fig3} \noindent
Horizontal Li$|$Cu metal junctions have spatial inhomogeneity linked to
galvanic corrosion.  (a) No oxide; (b) with Li$_2$O (111) film on top
(``oxide 1''); (c) with additional Li inserted below Li$_2$O film on the side
of Cu (``oxide 2'').  ${\cal V}_e$=2.14~V, 0.85~V, and 0.01~V, respectively.
Cu, Li, Li (oxide), and O atoms are depicted as yellow, purple, blue, and red
spheres.  (d)-(f) Local electrostatic potential integrated over the $y$
direction (${\phi}(x)$) for panels a-c, multiplied by $-|e|$.  More negative
values favor electron ejection from the Li metal, and the absolute values of
the vertical axis are arbitary.  In (d), red, green, and blue $\phi(x)$ are
computed with $z$-dimensions of 50, 40, and 28~\AA.  In panels (e)-(f), red,
green, and blue curves are computed with $z$-dimensions of 60, 50, and 40~\AA.  
Red and green almost coincide.  \color{black} We stress that $\phi(x)$ is
not a measurable quantity. \color{black} (g)-(h) Open circuit voltages measured
during Li self-discharge; the two curves depict different amount of partially
plated lithium (see S.I. Sec.~S2).  
}
\end{figure}

%\vspace*{0.1in}
\subsection {\it {\bf Cu$|$Li Junction: Work Function}}
%\noindent {\it {\bf Cu$|$Li Junction: Work Function}}

The above discussion lays the ground work for examining DOL electrochemical 
reduction on the Li$|$Cu junction.  Here we report the computational results
and defer discussions of relevance to corrosion to the next subsection.
First we distinguish two types of junctions.  Fig.~\ref{fig2}c depicts a
layer of Li (001) covering Cu (001).  This model mimics complete Li-plating
on a Cu current collector.  In contrast, Fig.~\ref{fig3}a depicts a junction
between Li and Cu metal in the $x$ direction.  It mimics partial,
$\sim$50~\%, Li plating on a Cu current collector.  

For the Fig.~\ref{fig2}c configuration, the work functions on the Li and Cu
sides of the simulation cells are 3.01~eV and 4.48~eV, which translate into
${\cal V}_e$=1.63~V and ${\cal V}_e$=3.11~V, respectively.  These values are
similar to those of pure Li and Cu metals with such facets; the $<$0.1~eV
discrepancy with published results\cite{gb} is due to the thinness of the
slabs and the small strain on the Cu, lattice matched to Li.  The similarity
with bare Li and Cu is in this case expected because electrostatic screening
length is on
the order of Angstroms within metals, and the effect of the junction should be
screened out.  The existence of two vacuum level is a DFT artifact,\cite{dipole}
and should be interpreted as follows.  The exposed Li and Cu surfaces are
electrically disconnected; they can be tuned to a common ${\cal V}_e$ via
modifying their independent EDL's.  Fig.~\ref{fig2}d further depicts a Li$_2$O
(111) slab coating the Li (001) surface on Cu(111).  This system exhibits
${\cal V}_e$=0.05~V, similar to the value for the Li (001)/Li$_2$O (111)
interface in the absence of the Cu slab (Fig.~\ref{fig2}a, without DOL).  

Next we consider the horizontal Li$|$Cu junction (Fig.~\ref{fig3}).  In
ground state DFT, a metal should exhibit a single $E_{\rm F}$ and a unique 
${\cal V}_e$.  Nevertheless, spatial heterogeneity is manifested in the local
electrostatic potential ${\phi}(x)$ in Fig.~\ref{fig3}d-f, \color{black} which
is a qualitative guide, not a measurable quantity.  \color{black} For the
Fig.~\ref{fig3}a configuration, $-|e|{\phi}(x)$, which suggests how favored
local $e^-$ ejection is,  has a maximum in the Cu region and a
mininum in the Li region that differ by \color{black} $\sim$1.26~eV \color{black}
(Fig.~\ref{fig3}d).  This should translate into a $\sim$1.26~V ``local
potential'' difference \color{black} within 3-6~\AA\, of the surface where
instantaneous \color{black} electrochemical reactions are
concerned.  If the two metallic regions were completely disconnected in the
$x$ direction, we expect $-|e|{\phi}(x)$ would be related to pure Li and pure
Cu vacuum level DFT/PBE values, which are 3.01~eV and 4.88~eV (see above),
yielding a 1.87~eV difference.  $e^-$ transfer from Li to Cu has evidently
reduced this ${\phi}(x)$ difference, but the Li side of the Li$|$Cu junction
\color{black} remains more electronegative than the Cu side within a few \AA\,
of the metal surface.  

Note that the 1.26~eV is obtained using a simulation
cell with $z$-dimensions of 40~\AA\, and 50~\AA.  The default 28~\AA\,
$z$-dimension cell (Table~S1, S.I.), which has a smaller vacuum region, 
yields a larger peak-to-value difference of 1.40~eV.  
The convergence of $\phi(x)$ is thus slower than the work function itself,
which is well converged with a $z$-dimension of 28~\AA. \color{black}

An uncoated electrode is unrealistic in battery settings.  Next, we put
a $\sim$10~\AA-thick Li$_2$O layer on the metal and optimize the atomic
configuration (Fig.~\ref{fig3}b, henceforth the ``oxide 1'' model).  We
predict ${\cal V}_e$=0.85~V for this system.  Joining Cu with Li has
therefore raised ${\cal V}_e$ by 0.80~V relative to Fig.~\ref{fig2}a.
At ${\cal V}_e$$>$0~V, the Li metal is metastable; however, the simulation
cell lacks a liquid electrolyte that can accommodate Li$^+$ or a simulation
time scale needed for dissolution to occur, and the Li slab remains.  The
difference between the Cu-maximum and the Li-minimum is reduced to \color{black}
0.39~eV \color{black} in the simulation cells with the largest vacuum
region (Fig.~\ref{fig3}e), but remains significant.  
%This $-|e|{\phi}(x)$ difference depends on the contact potentials which should
%not be strongly affected by oxide thicknesses.\cite{pccp} 

Finally, we find that adding \color{black} 14 \color{black} additional
``interlayer'' Li atoms beneath the Li$_2$O film on the Cu side (Table~S3
in the S.I.), with each Li directly coordinated to a O~anion at the interface,
lowers ${\cal V}_e$ to  0.01~V,\cite{gb} which is almost at $i$-equilibrium
(${\cal V}_e$=${\cal V}_i$=0.0~V, Fig.~\ref{fig3}c).  (Again, ${\cal V}_e$ is
well converged with respect to the $z$-dimension of the the simulation cell;
increasing the cell size by 14~\AA\, reduces ${\cal V}_e$ only by 0.03~V;
henceforth we report the 0.01~V value.)  This model is
referred to as ``oxide 2.'' We stress that, for this model, we have managed
to change ${\cal V}_e$ significantly by only altering the contact
potential (SI~Sec.~S11);\cite{gb,pccp,marks} this solid-state degree of freedom
at Li-battery interfaces, arising from the perhaps unique mobility of Li$^+$,
is missing at the interface of uncoated electrodes,\cite{otani,filhol} but
cannot be neglected in batteries.  The average binding energy of the
14~added Li atoms, offset by the Li metal cohesive energy, is a favorable
-0.57~eV, consistent with ${\cal V}_i$=0.57~V assuming that these are the most
favorable available Li-sites.  In other words, Li insertion into these
interface sites is thermodynamically favorable at or below 0.57~V, although we
have not examined all possible Li configurations, and the estimated
${\cal V}_i$ is approximate. \color{black} Creation
of this interlayer should occur when the instantaneous voltage is
${\cal V}_e$=0.01~V (Fig.~\ref{fig3}c) when the interlayer is
thermodynamically favored, but not at 0.85~V (Fig.~\ref{fig3}b).  Strictly
speaking, adding 18~Li at the interface is found to be most consistent
with ${\cal V}_i$=0.0~V, but this yields ${\cal V}_e$=-0.16~V (SI Sec.~S10).
Here we neglect this slight discrepancy in ${\cal V}_e$.  Oxide 1~and~2 models
are approximate, self-consistent depictions of the system at two different
applied voltages at time scales before Li from the Li metal can diffuse;
\color{black} oxide~2 conforms to our definition of $i$-equilibrium.  
Interlayer Li atoms have significantly less effects on the Li side.\cite{gb}
After adding these Li atoms to the Cu side, the maximum variation in
$-|e|{\phi}(x)$ is further lowered to \color{black} 0.16~eV, \color{black}
(Fig.~\ref{fig3}f), but still persists.

Fig.~\ref{fig3}d-f emphasize the existence of spatial electrostatic 
heterogeneity on metal electrode surfaces.  
Local voltage variation of up to 0.3~V during charge has
been measured for graphite anodes.\cite{mukherjee} This variation has been
entirely attributed to kinetic limitations/ohmic losses; our predictions
suggest that structural origins may also need to be explored. 

\color{black}
%\vspace*{0.1in}
\subsection {\it {\bf Interpretations of Cu$|$Li Junction Work Functions}}
%\noindent {\it {\bf Interpretations of Cu$|$Li Junction Work Functions}}

Our calculations show that a 100\% Li$_2$O(111)-coated Li(001) slab 
(Fig.~\ref{fig2}a without a DOL) exhibits a similar work function as the same
composite slab deposited on Cu (001) (Fig.~\ref{fig2}d).  This suggests that,
on copper current collectors completely covered with plated Li metal
(Fig.~\ref{fig2}c-d), and therefore without a electrolyte/Li/Cu 3-phase
junction, the Li/Cu contact has no effect on electron transfer.  Hence
galvanic corrosion should not occur.  This inference is consistent with
Ref.~\onlinecite{galvanic2} and~\onlinecite{merrill}.  

Next we consider the horizontal Li$|$Cu slabs.  The ``galvanic corrosion''
concept is arguably most relevant in field deployment environments (e.g.,
on ships\cite{jced}) where the overall metal voltage (${\cal V}_e$) is not
controlled.  In such conditions, we expect that creating a junction of two
metals with different electronegativities raises ${\cal V}_e$ above that of
the more electronegative metal alone, at least prior to any induced
electrochemical reaction that has occurred and has led to changes in 
${\cal V}_e$.\cite{kelly}  This is indeed the situation for the oxide~1 model,
where the three phase junction between Li$_2$O, Li, and the more
electropositive Cu raises the voltage from the ${\cal V}_e$=0.05~V value at
the pure Li$_2$O/Li interface (Fig.~\ref{fig2}) to 0.85~V at the mixed
interface (Fig.~\ref{fig3}b). Li dissolution, and pitting through the oxide,
would be accelerated at this higher ${\cal V}_e$.  This scenario would be
consistent with standard galvanic corrosion.\cite{jced}  Note that Li$^+$
dissolution requires an $e^-$ acceptor.  In the next section we show that DOL
reduction is unfavorable at ${\cal V}_e$=0.85~V for this model; however, other
electrolyte components, e.g., the anion, exhibit higher reduction potentials
and can accept $e^-$.
This model bears similarities with Al corrosion. metal ion transport in the
Al$_2$O$_3$ passivating film is slow prior to pitting.\cite{corrosion} The
onset potential for pitting depends on pH and salt concentration but is
generally at values much higher than the standard Al plating
voltage.\cite{pittingvoltage} A large overpotential vs.~metal-stripping is
readily attainable in table-top electrochemical settings.  At this onset,
the passivating oxide is ruptured and Al$^{3+}$ can readily dissolve.  

However, we argue that Fig.~\ref{fig3}e is less relevant than Fig.~\ref{fig3}f
in batteries.  Li$^+$ diffuses much more readily in the SEI passivating film
than Al$^{3+}$ in its oxide without the need of first rupturing the SEI film.
The lowering of kinetic constraint means that the anode should be near its
equilibrium potential.  Indeed, our measurements show that the Li-stripping
occurs without significant over-voltage (${\cal V}_e$$\sim$0~V)
(Fig.~\ref{fig3}g-h), quantitatively similar to ``oxide~2,'' and is not at
0.85~V.  This should be true in recent battery experiments as
well.\cite{galvanic1,galvanic2,merrill} 

In that sense, oxide~1 (Fig.~\ref{fig4}b) is a hypothetical, standard
``galvanic corrosion'' scenario that assumes that Li$^+$ motion in the SEI is as
slow as Al$^{3+}$ in alumina.  Oxide~2, where ${\cal V}_e$$\sim$0~V at room
temperature, is more relevant to battery experiments (Fig.~\ref{fig3}g-h).
Even in this case, the DFT-predicted work function and local electrostatic
potential $-|e|{\phi}(x)$ remain {\it lower} on the Li side of the Li$|$Cu
metal junction (Fig.~\ref{fig3}f).  The expected transfer of $e^-$ from the Li
side to the Cu side due to their difference in electronegativity reduces the
difference in $-|e|{\phi}(x)$ arising from the pure Li and Cu metal work
function difference, but does not eliminate or reverse it.  For a Cu$|$Al
metal junction, finite element simulations have also predicted local
voltage increase on the more electropositive Cu side.\cite{haque}  This is
qualitatively simillar to Fig.~\ref{fig3}d-f, although the variation is much
smaller, likely because an electric current exists in the model of
Ref.~\onlinecite{haque}, and ohmic effects cannot be neglected.  A more direct
connection between metal corrosion (``oxide~1''), and SEI at a substantial
potential ($\sim$0.85~V), may however be made using passive anodes without
Li content.\cite{maureen}  

SEI forms preferentially at lower voltages and metal dissolution into cations
(e.g., via pitting) preferentially occurs at higher voltages.  In the oxide~2
model (Fig.~\ref{fig4}c), the $-|e|{\phi}(x)$ variation in Fig.~\ref{fig3}f
suggests that, if the SEI thickness are the same on both Cu and Li sides, the
Li side is at a $<$0.0~V potential.  This suggests that the partially plated
Li-regions in recent experimental work\cite{galvanic1,galvanic2,merrill}
suppress Li dissolution via destruction of the passivating film
(pitting-like corrosion), while electrolyte electrochemical reduction is
enhanced.  The Li-plated region is preferentially the cathode
(electrochemically reducing electrolyte).  \color{black} ``Pitting'' does not
occur in the Li-plated region; instead Li$^+$ dissolution may occur at the
Li$|$Cu junction or on the Cu-side where the final SEI is thinner. \color{black}
This would be a ``cathodic'' reaction driven corrosion -- at
least at the thin SEI thickness in our model -- which is fundamentally different
from standard anodic-limited local Al or steel corrosion initiation mechanisms;
it may be more similar to Mg corrosion in water where the passivating film is
less stable and hydrogen reduction reaction occurs more readily.\cite{mgwater}
This appears to contradict
the proposed mechanism (Fig.~4 schematic) in Ref.~\onlinecite{galvanic1}.
However, the Ref.~\onlinecite{galvanic1} analyses are performed after the SEI
is fully formed, at which point the Li-side SEI is thicker than that on the
Cu-side (unlike in our models) and SEI formation is more kinetically limited.
Our argument concerning $-|e| {\phi}(x)$ does not need to be altered to make
this connection to the experimental work.  SEI formation reactions will be
explicitly examined in the next sections.  

Thus we caution against a too-literal identification of traditional
corrosion/galvanic corrosion concept in aqueous electrolytes with Li
self-discharge.  Future local voltage measurements in Li-plated and non-plated
Cu current collectors, e.g., by using scanning Kelvin probe force microscopy,
will be valuable to confirm this point.

Note that the experimental voltage (Fig.~\ref{fig3}g-h) is slightly but
measurably higher for a sample where Li has been plated on Cu at a very 
low capacity than when plated at higher capcity, especially at longer
times.  See the S.I. (Sec.~S2) for details.  The $<$0.01~V difference at
early times cannot be resolved in current DFT calculations.  At the lower
Li capacity, we expect the large Cu surface area relative to the
amount of plated Li to lead to rampant self-discharge. This suggests
that when significant self-discharge occurs, the open circuit voltage is
higher than the Li stripping potential of 0.0~V, which accelerates
Li$^+$ dissolution.  This is in {\it qualitative} agreement with the oxide~1
model, although we stress that ``oxide~1'' is not meant to be physical, and
the experimentally measured voltage is not as high as the 0.85~V.  The
difference in magnitude may be due to many factors, such as the effective
time-scale difference, S.I., allowing change of Li configuration from oxide~1
to oxide~2.

\color{black}

%\vspace*{0.1in}
\subsection{\it {\bf Cu$|$Li Junction: Electrochemical Reduction of Solvent}}
%\noindent {\it {\bf Cu$|$Li Junction: Solvent Electrochemical Reduction}}

\begin{figure}
%\centerline{\hbox{ (a) \epsfxsize=2.70in \epsfbox{fig4a.ps} 
%                   \epsfxsize=2.70in \epsfbox{fig4b.ps} (b) }} \caption[]
\centerline{\hbox{  \epsfxsize=5.50in \epsfbox{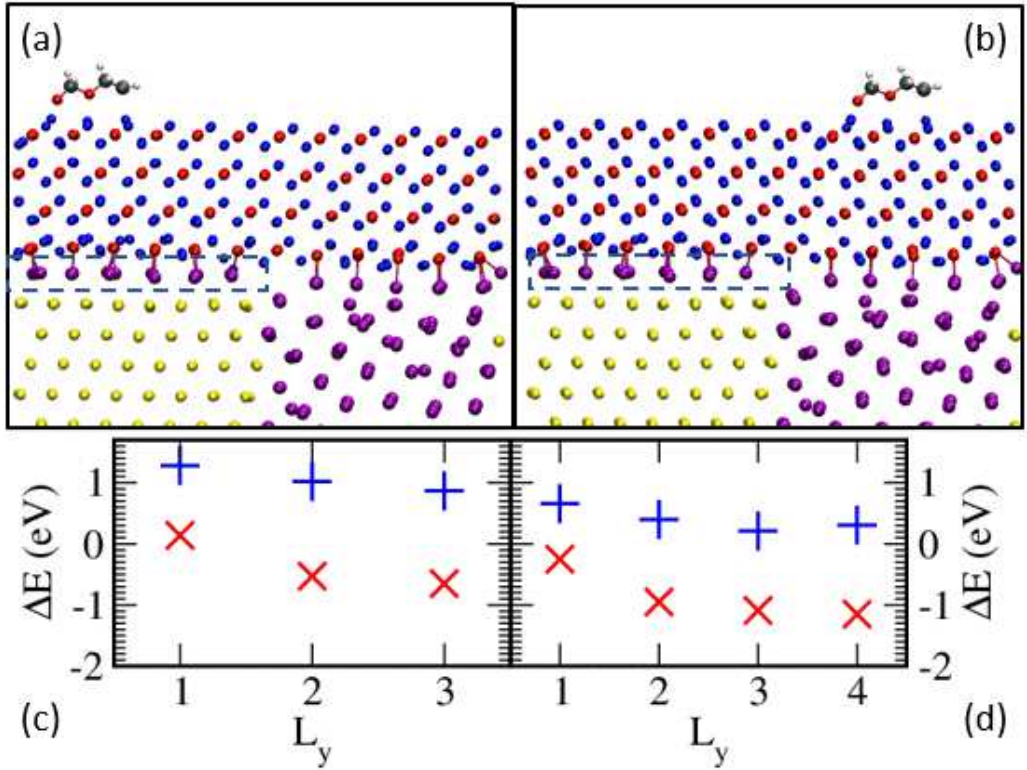} }}
\caption[]
{\label{fig4} \noindent
(a)-(b) correspond to Fig.~\ref{fig3}c (oxide~2), but with a decomposed DOL
molecule on the Li and Cu sides, respectively.  ``Interlayer'' Li on
the Cu-side are highlighted in a box.  (c)-(d) Energy cost ($\Delta E$)
associated with decomposing DOL molecules on the Li and Cu side as
the system size ($L_y$) increases, respectively.  Red and blue correspond to 
${\cal V}_e$=0.01~V and 0.85~V (with and without interlayer Li, i.e.,
oxide~2 and oxide~1), respectively; DOL reduction decomposition is always more
favored on the Li-side, and is thermodynamically favorable/unfavorable at
0.01~V/0.85~V.
}
\end{figure}

Local electrostatic potential variation should be screened by low dielectric
organic SEI far from electrode surfaces, but reduction reactions occurring
right at the inoroganic SEI surface should be affected by the variation in
${\phi}(x)$ (Fig.~\ref{fig3}d-f).  We next demonstrate the consequences
of $-|e|{\phi}(x)$ heterogeneity, and of overpotential, on interfacial
stability.  We compute DOL reductive decomposition energetics on ``oxide 2''
(Fig.~\ref{fig3}c) surfaces where the overall ${\cal V}_e$ is at the Li-plating
voltage.  Fig.~\ref{fig4}a-b depict a decomposed DOL on the Li and the Cu
sides of the oxide~2 model, at the peak and valley of the $-|e|{\phi}(x)$
profile (Fig.~\ref{fig3}f), respectively.  Two-electron DOL decomposition is
more favorable by
\color{black}
$\Delta E$=-1.15~eV on the Li-side versus -0.65~eV on the Cu-side
\color{black}
in the largest supercells examined (Table~\ref{table1}).  The $\Delta E$
difference between the two sides is 0.50~eV, reasonably close in magnitude
than 2$|e|$ times the \color{black} 0.16~eV \color{black}
variation in $-|e|{\phi}(x)$ (Fig.~\ref{fig3}f); the latter
is meant as a qualitative guide for explicit $e^-$ transfer reactions.
%On the Li-side, the $\Delta E$ is more negative than the -1.72~eV value
%associated with Fig.~\ref{fig2}b in the absence of Cu, because DOL
%electrochemical decomposition in Fig.~\ref{fig4}b occurs at lower than
%the 0.05~V condition of Fig.~\ref{fig2}a due to the ${\phi}(x)$ variation.

In contrast, the oxide~1 model exhibits a significantly higher overall
(${\cal V}_e$=0.85~V), which should impede DOL reductive decomposition that
leads to SEI-formation.  Indeed, $\Delta E$=+0.31~eV and +0.87~eV on the Li- and
Cu-sides, respectively in the largest cells considered (Table~\ref{table1}).
The $\Delta E$ discrepancy between the Li and Cu sides on the oxide~1 model
is qualitatively consistent with the larger $-|e|{\phi}(x)$ variation
(Fig.~\ref{fig3}e), but again ${\phi}(x)$ does not quantitatively describes
the $\Delta E$ difference.

The reaction barriers $\Delta E^*$ for the forward reaction are irrelevant
on ``oxide~1'' surfaces because the reactions are endothermic at the
${\cal V}_e$=0.85~V voltage there, and DOL decomposition should not occur.
However, the reverse reaction, namely reconstitution of intact DOL molecules,
is energetically downhill, and the reverse $\Delta E^*$ are lower than 1.0~eV
(Table~\ref{table1}).  This implies reaction time scales of at most 1~hour
when using standard kinetic prefactors in Arrhenius rate estimates at T=300~K. 
$\Delta E^*$ for DOL decomposition in the largest simulation cell used
is 0.80~eV (Table~\ref{table1}), which indicates that the reaction is fast
relative to battery cycling ($\sim$1-hour) time scales.
At the transition state, the C-O bond being broken has a length of 2.11~\AA\,
in the oxide~2 model but is a much smaller 1.98~\AA\, in the oxide~1 model,
suggesting that differences exist between DOL decomposition
mechanisms at different ${\cal V}_e$.
At ${\cal V}_e$=-0.16~V, $\Delta E$ is significantly more negative than
in ``oxide 2,'' and $\Delta E^*$ is even lower (S.I.~Sec.~S10); the
DOL decomposition energetics may be non-linearly dependent on ${\cal V}_e$.

\color{black} $\Delta E$ are not strongly dependent on the size of the vacuum
region, unlike $\phi(x)$ (Fig.~\ref{fig3}).  Increasing the cell size from
$z$=40~\AA\, used in Fig.~\ref{fig4} to $z$=50~\AA\, on the Cu side of
the oxide~1 model changes $\Delta E$ by at most 0.06~eV; the variations of
$\Delta E$ between the Li- and Cu-sides are converged to within a few meV's.
\color{black}

These $\Delta E$ predictions associated with Fig.~\ref{fig3}b-c should
definitively dispel the much-quoted misconception that voltages in DFT
simulation cells are solely determined by ${\cal V}_i$, which are the Li
insertion costs relative to Li metal cohesive energy.\cite{voltage_prb}
Li metal is present in both oxide~1 and oxide~2, but the $\Delta E$
associated with DOL decompositon due to cross-SEI-film $e^-$ transfer,
which is a prototype voltage-dependent SEI formation reaction, varies
significantly.  $\Delta E$ even depends on the side of the simulation cell
in which the reaction occurs.  An equally incorrect corollary assumption is
that the DFT simulation cell is always at electrochemical equilibrium 
(${\cal V}_e$=${\cal V}_i$=0.0~V vs.~Li$^+$/Li(s)) if lithium metal exists
in the cell.  While ${\cal V}_i$ is the correct equilibrium definition, it
does not account for the possibility of instantaneous overpotential in DFT, or
experimental, settings (S.I.~Sec.~S5).  \color{black} Unless ${\cal V}_e$ is
controlled by tuning the interfacial structure, it is very easy have
accidental DFT overpotential that lead to qualitatively wrong conclusions
(e.g., by incorrectly assuming that ``oxide~1'' represents a system is at
$i$-equilibrium and that DOL does not decompose on Li anode surfaces at
$i$-equilibrium, Fig.~\ref{fig4}a,c).  This emphasizes the point that all DFT
modeling work on battery interfaces should calculate and report ${\cal V}_e$.
\color{black}
When long-range $e^-$ transfer-induced SEI formation reaction occurs, the Li
metal is not in contact with the species being electrochemically reduced; the
surface-film coated electrode in effect acts like a passive electrode at
short times (S.I.~Sec.~S5) just like in supercapacitors, and the electronic
voltage ${\cal V}_e$ used in supercapacitors and related disciplines is the
correct definition. It is only by considering ${\cal V}_e$ that the
predictions in Fig.~\ref{fig4} can be rationalized.  

\begin{table}\centering
\begin{tabular}{l l l l r r r} \hline
system & side & size & $\Delta E$ & $\Delta E^*$ 
		& ${\cal V}_e$ &  $\Delta {\cal V}_e$ \\ \hline
Li/Li$_2$O  & NA & 1$\times$1  & -0.85 eV & +0.83 eV & 0.09 V & 2.08 V \\
Li/Li$_2$O  & NA & 2$\times$1  & -1.55 eV & +0.72 eV & 0.12 V & 0.89 V \\
Li/Li$_2$O  & NA & 2$\times$2  & -1.72 eV & +0.67 eV & 0.01 V & 0.14 V \\ \hline
Li+Cu/Li$_2$O  & Li$^a$ &  1$\times$1  & +0.67 eV & 0.75 eV & 0.77 V & 1.25 V \\
Li+Cu/Li$_2$O  & Li$^a$ &  1$\times$2  & +0.40 eV & 0.91 eV & 0.83 V & 0.78 V \\
Li+Cu/Li$_2$O  & Li$^a$ &  1$\times$3  & +0.21 eV & NA & 0.79 V & 0.54 V \\ 
Li+Cu/Li$_2$O  & Li$^a$ &  1$\times$4  & +0.31 eV & NA & 0.79 V & 0.42 V \\ 
									\hline
Li+Cu/Li$_2$O  & Cu$^a$ &  1$\times$1  & +1.26 eV & NA & 0.75 V & 0.99 V \\
Li+Cu/Li$_2$O  & Cu$^a$ &  1$\times$2  & +1.02 eV & NA & 0.82 V & 0.59 V \\
Li+Cu/Li$_2$O  & Cu$^a$ &  1$\times$3  & +0.87 eV & NA & 0.78 V & 0.43 V \\ 
									\hline
Li+Cu/Li$_2$O  & Li$^b$ &  1$\times$1  & -0.25 eV & 0.97 & -.04 V & 1.60 V\\
Li+Cu/Li$_2$O  & Li$^b$ &  1$\times$2  & -1.09 eV & 0.80 & -.05 V & 0.95 V \\
Li+Cu/Li$_2$O  & Li$^b$ &  1$\times$3  & -1.09 eV & NA & -.00 V & 0.65 V \\ 
Li+Cu/Li$_2$O  & Li$^b$ &  1$\times$4  & -1.15 eV & NA & -.03 V & 0.48 V \\ 
									\hline
Li+Cu/Li$_2$O  & Cu$^b$ &  1$\times$1  & +0.14 eV & 1.12 & -.04 V & 1.46 V \\
Li+Cu/Li$_2$O  & Cu$^b$ &  1$\times$2  & -0.53 eV & NA & -.05 V & 0.90 V \\
Li+Cu/Li$_2$O  & Cu$^b$ &  1$\times$3  & -0.65 eV & NA & -.00 V & 0.63 V \\ 
									\hline
LiH	       &  NA   &   1$\times$1  & -1.17 eV & +2.13 eV & NA & NA \\ 
\hline
Li/LiH	       &  NA   &   1$\times$1  & -0.74 eV & +1.04 eV & 0.63 V & 1.10 V\\
Li/LiH	       &  NA   &   $2^{0.5}$$\times$$2^{0.5}$  & -0.99 eV & +0.98~eV &
					+0.65 V & 0.56 V \\ 
Li/LiH	       &  NA   &   2$\times$2  & -0.89 eV & 0.93 eV & 0.67 V & 0.28 V\\
	\hline
\end{tabular}
\caption[]
{\label{table1} \noindent
DOL decomposition energy ($\Delta E$), barrier ($\Delta E^*$),
electronic voltage before DOL decomposes (${\cal V}_e$), and change in
electronic voltage after DOL docomposes ($\Delta {\cal V}_e$)  as functions of
system, system supercell size, and (in the case of Li+Cu electrodes) whether
the DOL is on the Li or Cu side.  Superscripts $a$ and $b$ refer to the
``oxide 1'' and ``oxide 2'' models (see text).  Supercell sizes are multiples
of the baseline unit cell in the lateral dimensions
(S.I. Table~S3).  $\Delta {\cal V}_e$ should converge to zero in
the limit of infinite size.  The ``LiH'' slab has no metallic anode and
${\cal V}_e$ is considered not well defined.  For oxide~1 models, $\Delta E^*$
refer to DOL reconstitution (not decomposition) reactions.
}
\end{table}

%\vspace*{0.1in}

\color{black}

\subsection{\it {\bf Interpretations of Solvent Reduction at Cu$|$Li Junction}}
%\noindent {\it {\bf Interpretations of Solvent Reduction at Cu$|$Li Junction}}
%\vspace*{0.1in}

Our calculations suggest that the main ``galvanic'' effect of the Li$|$Cu
junction is to enhance SEI formation on Li-plated regions over
Cu regions.  This is a cathodic reaction-driven process, distinct from
traditional Al/steel pitting corrosion.

We stress that SEI films are thin ($\sim$10~\AA) in our models, and are not yet
completely passivating, unlike in experiments where the inorganic SEI on the
Li-side is completed and is reported not increase in thickness during
subsequent Li dissolution in cryo-TEM measurements, at least for some
cycles.\cite{galvanic1} In our models, more favorable DOL decomposition
enthalpy ($\Delta E$'s) are predicted on the Li side of Cu$|$Li junctions in
both oxide~1 and oxide~2 at the same Li$_2$O thickness.  This enhanced
tendency on Li, when the Li$_2$O thickness is the same on both sides,
is consistent with the thicker ($\sim$15~nm) SEI ultimately formed on Li
compared to that on Cu ($\sim$5-8~nm).\cite{galvanic1}  
Measurement of SEI thickness on the Cu side as Li self-discharge proceeds
would be valuable in future studies.  Although we have focused on initial
DOL decomposition reactions and have not explored ultimate SEI products,
the different values of $\Delta E$ and $\Delta E^*$ predicted on the Li and
Cu sides may be consistent with the different SEI compositions in the two
types of regions.\cite{galvanic1}  Indeed, the SEI formation potential has
been reported to affect SEI composition and properties on other
substrates.\cite{evolve3}  See the S.I. (Sec.~S1) for further XPS SEI analysis.

As discussed above, the ``oxide~1'' model is at significant overpotential
(0.85~V) vs.~Li-stripping.  Such high voltages may be observed only after all
active Li has been removed from the anode (Fig.~\ref{fig3}g-h).  The reverse
reaction barriers
($\Delta E^*$) calculated for that model are likely underestimated because
other decomposed DOL fragments and other Li$^+$ cations, not included in
the model, may bind to and stabilized the isolated, electrochemically reduced
DOL in the equivalent of Fig.~\ref{fig4}a-b for oxide~1 (without interlayer
Li).  However, our calculations raise the possibility SEI evolution at voltages
higher than Li-stripping may involve reversal of SEI formation reactions.  
Note also that there are differences in the electrolytes used in this work
and in Refs.~\onlinecite{galvanic1,galvanic2,merrill}.

\color{black}

%\vspace*{0.1in}
\subsection{\it {\bf LiH and DOL Reactions: Chemical vs.~Electrochemical
Self-discharge}}
%\noindent {\it {\bf LiH and DOL Reactions: Chemical vs.~Electrochemical
%Degradation}}

LiH has recently reported to be an SEI component,\cite{lih1,lih2} although
this finding has been disputed.\cite{lih3}  Next we consider LiH in the
SEI.  H$^-$ is in its lowest possible charge state and, like
other inorganic SEI components Li$_2$O and LiF, LiH should be thermodynamically
stable against Li metal.  But unlike those components, LiH is a strong reducing
agent that should react chemically with organic SEI and electrolyte components,
even in the absence of an $e^-$ source.  Thus, apart from the inherent
interest in LiH within the SEI, this material serves as a model system to
differentiate chemical versus electrochemical reactions with the electrolyte
during SEI evolution, late-stage SEI growth, or self-discharge.

\begin{figure}
 \centerline{\hbox{ (a) \epsfxsize=2.00in \epsfbox{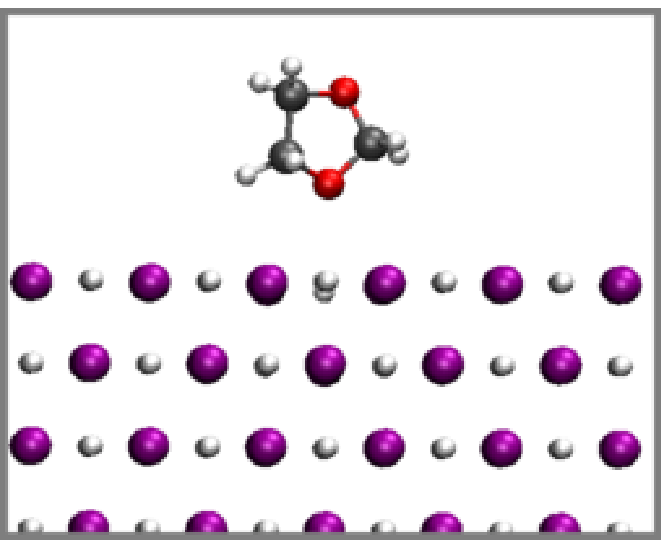} 
                    \epsfxsize=2.00in \epsfbox{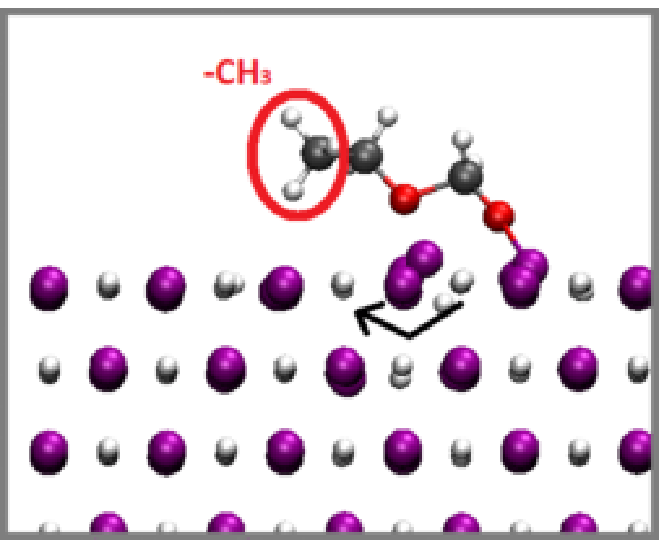} (b) }}
 \centerline{\hbox{ \epsfxsize=1.50in \epsfbox{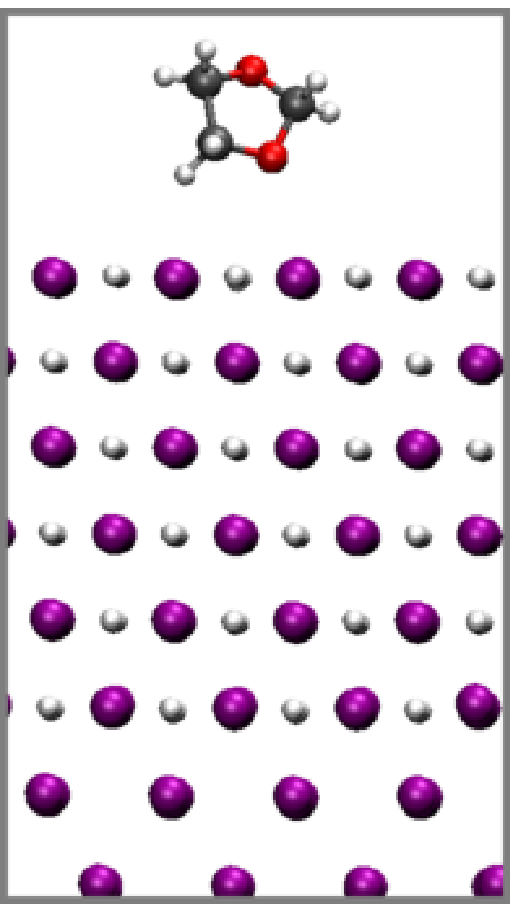}
                    \epsfxsize=1.50in \epsfbox{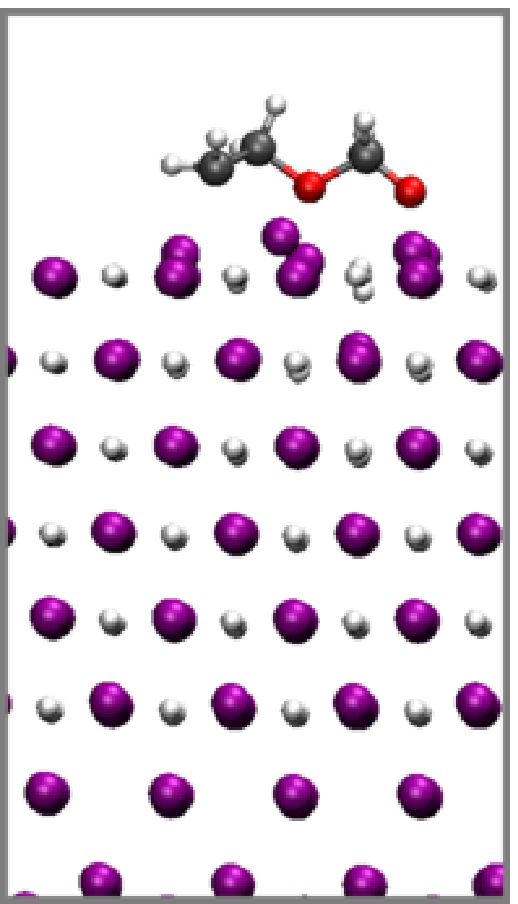} 
                    \epsfxsize=1.50in \epsfbox{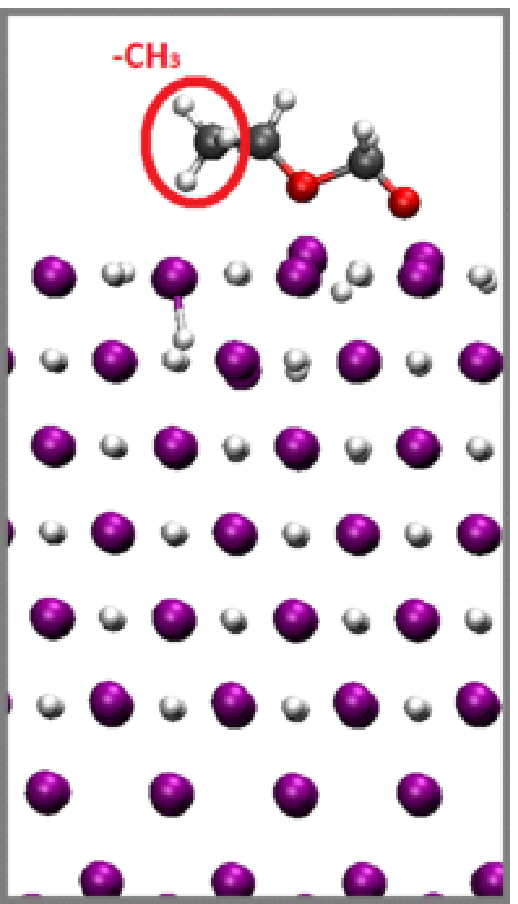} }}
 \centerline{\hbox{ (c) \hspace*{1.2in} (d) \hspace*{1.2in} (e) }}
%\centerline{\hbox{  \epsfxsize=5.50in \epsfbox{fig5.ps} }}
\caption[]
{\label{fig5} \noindent
(a)-(b) Intact and decomposed DOL molecule on LiH (001) surface, no Li metal
anode (chemical reaction).  In (b), a surface H$^-$ has been transferred to
the terminal CH$_2$ group; $\Delta E$=-1.17~eV and $\Delta E^*$=+2.13~eV
relative to (a).  The black arrow depicts a further reaction pathway involving
H$^-$ diffusion, and red circles indicate CH$_3$ groups (if any).  Unreacted DOL
does not contain CH$_3$ groups.  (c)-(e) Electrochemical or chemical
reaction.  (c) Intact DOL on LiH (001)-coated Li metal surface.  (d) Decomposed
DOL via breaking of C-O bond; $\Delta E$=-0.89~eV and $\Delta E^*$=0.67~eV
relative to (c).  (e) Decomposed DOL via abstraction of H from LiH surface;
the leftmost C-atom has 3 covalently bonded H.  $\Delta E$=-1.24~eV relative
to (c).  Li, C, O, H are purple, grey, red, and white.
}
\end{figure}

Fig.~\ref{fig5}a-b depict intact and decomposed DOL configurations on the
LiH (001) surface.  These simulation cells omit metallic electrodes which
are $e^-$ sources, and only chemical reactions are possible.  In panel (b),
a H$_2$C-O bond is broken, and one of the H$^-$ on the LiH surface is now
covalently bonded to the exposed CH$_2$ group.
$\Delta E$=-1.17~eV (Table~\ref{table1}), indicating that the LiH surface
reacts exothermically with DOL.  An even more exothermic configuration, with
$\Delta E$=-2.63~eV, can be found if the H$^-$ surface ``hole'' is filled by
another H$^-$ from the LiH (marked by an arrow in Fig.~\ref{fig5}b).  We
expect that other organic molecules react exothermically with LiH as well.

However, SEI formation typically occurs at room temperature and the reactions
are governed by kinetics, not just thermodynamics. Going from panel (a) to (b),
the kinetic barrier $\Delta E^*$=+2.13~eV is unexpectedly large, and the
reaction cannot occur at T=300~K within battery operation time scales.  As
this simulation cell lacks a metal anode we do not report ${\cal V}_e$.  No
long-range charge transfer occurs here, and system size effects are expected
to be small.
%For a comparison, we predict that  $\Delta E^*$ is +0.74~eV for DOL
%decomposition on Li (001) surfaces (SEI), which is at millisecond time scale. 

Next we add lithium metal to the backside of the LiH slab.  The LiH (001)/Li
(001) interface exhibits ${\cal V}_e$=+0.63~V without an adsorbed, intact
DOL.  This is significantly higher than the Li-plating potential.  It
indicates that the Li$_2$O (111)/Li (001) and LiF (001)/Li (001) interfaces,
which both exhibit ${\cal V}_e$$\sim$0.0~V without the need to create surface
charges, are not universal features of Li/SEI interfaces.  \color{black} It also
emphasizes that extra effort is needed to keep DFT model systems at
$i$-equilibrium, non-overpotential conditions. \color{black}

Fig.~\ref{fig5}c-d depict an intact and electrochemically decomposed DOL on
the (001) surface of LiH-coated lithium metal, respectively.  ${\cal V}_e$ is
0.67~V for the intact DOL structure (Fig.~\ref{fig5}a), similar to the 0.63~V
computed without the DOL.  The decomposed DOL has the same structure as
that on Li$_2$O (111) (Fig.~\ref{fig2}b, Fig.~\ref{fig4}b)).  A C-O bond is
broken, but there is no H$^-$ transfer from LiH to the DOL, unlike
Fig.~\ref{fig5}b.  $\Delta E$ and $\Delta E^*$ are -1.32~eV and +1.04~eV for
this reaction, respectively, in the largest cell considered.  Since this
system is at an overpotential vs.~Li$^+$/Li(s), DOL electrochemical reduction
at Li-plating potential (0.0~V vs.~Li$^+$/Li(s)) should be even more favorable.
In contrast, Fig.~\ref{fig5}e depicts a reaction route which involves transfer
of a H$^-$ to the DOL just like Fig.~\ref{fig5}b.  $\Delta E$=-1.24~eV, very
similar to the -1.17~eV without Li metal (Fig.~\ref{fig5}b).  ${\cal V}_e$
is 0.67~V for the decomposed DOL, similar to that for Fig.~\ref{fig5}c.  The
lack of a change in ${\cal V}_e$ in a finite sized cell suggests that no $e^-$
is transferred, little change in dipole moment has occurred from the Li metal,
and therefore the reaction is chemical, not electrochemical, in nature.  As
expected, the presence of the Li metal does not strongly affect purely chemical
reactions.  We have not computed $\Delta E^*$ but it is likely to be similar to
the system without Li metal (Fig.~\ref{fig5}a-b), and makes the reaction
kinetically unfavorable.

For $\Delta E$ calculations on LiH film on Li metal slabs, the convergence
with system size is significantly faster than on Li$_2$O films
(Fig.~\ref{fig4}c-d).  The reason may be the smaller LiH band gap, which
translates into a larger dielectric constant; furthermore, the conduction
band edge is much closer to $E_{\rm F}$ than Li$_2$O-coated Li metal (S.I.
Sec.~S3), which also increases the polarizability of the entire system.
\color{black} We have not considered additional mechanistic steps, like
Li$^+$ dissolving from the Li metal and coordinating to the DOL molecule,
partly because the Fig.~\ref{fig5}d configuration is already sufficiently
favorable in terms of both energetics and kinetics; if adding more
mechanistic complexity further lowers $\Delta E^*$ or $\Delta E$, it would
not change our conclusion. \color{black}

These results indicate that in general, intermediate stage SEI growth, which
consumes the electrolyte, can occur via both electrochemical and chemical
routes.  For LiH, the chemical route has high reaction barriers while the
electrochemical route does not.  If LiH is present in electrically
disconnected regions of the SEI, it may have significant kinetic stability
against the organic solvent and organic SEI components, and may exhibit a
significant lifetime.  Furthermore, if LiH is continously generated by
electrolyte reaction with Li metal, the subsequent reaction of LiH with
the electrolyte also leads to a net consumption of Li, and therefore
self-discharge.  

%\vspace*{0.1in}
\subsection{\it {\bf Li$^+$ Vacancy at Li/LiAlO$_2$ Interface: 
Electric Field Effects}}
%\noindent {\it {\bf Li$^+$ Vacancy at Li/LiAlO$_2$ Interface: 
%Electric Field Effects}}

In this section we survey two other aSEI/Li interfaces examples, namely,
LiAlO$_2$ and LiI, where ${\cal V}_e$ are not naturally at 0.0~V
vs.~Li$^+$/Li(s), discuss how to shift ${\cal V}_e$ towards the Li-plating
potential, and examine the potential consequences to battery operations.
\color{black}
We also focus on the relative energy landscape related to Li$^+$ vacancy
diffusion and show that Li$^+$ diffusion towards/away from the interface
can become asymmetric.  We do not focus on the absolute Li$^+$ vacancy
formation energies; at a given temperature there will be a certain population
of such vacancies.  Our explicit interface models used herein go beyond
traditional semiconductor models\cite{freysoldt} and permit the treatment
of thin SEI films, surface charges, electric fields, and contact potentials.
\color{black}

\begin{figure}
\centerline{\hbox{ (a) \epsfxsize=1.50in \epsfbox{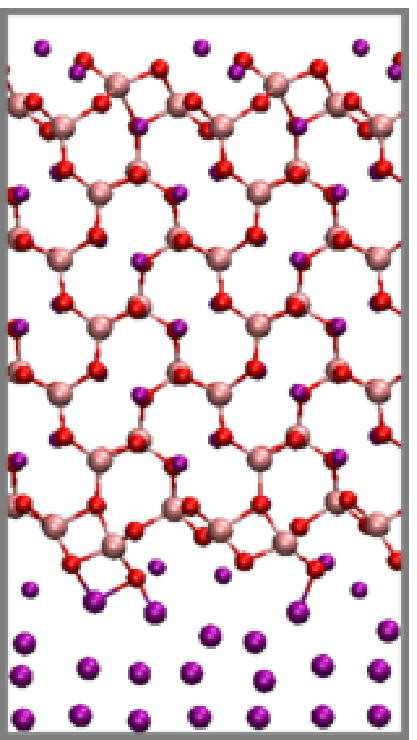} 
		   (b) \epsfxsize=2.80in \epsfbox{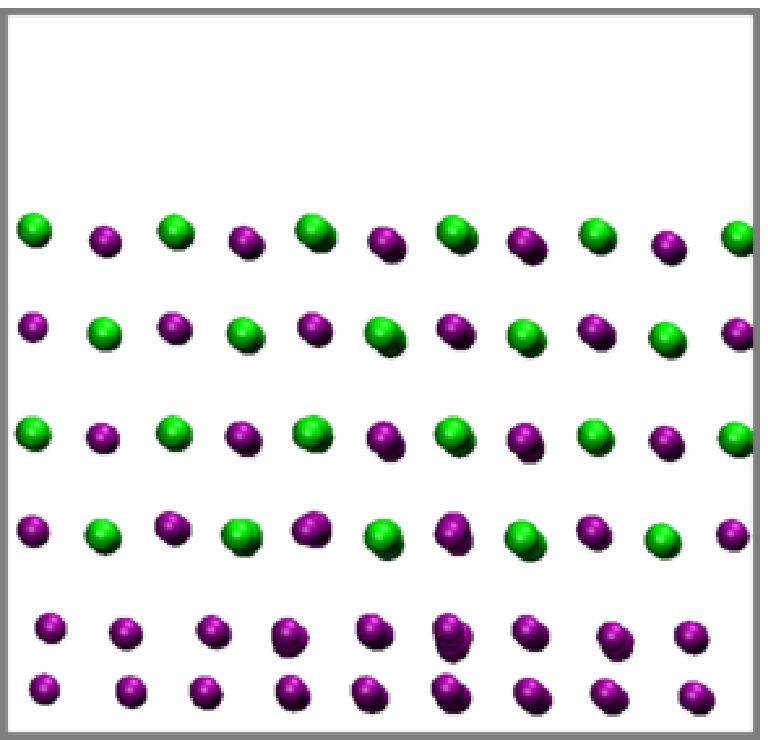} }}
\caption[]
{\label{fig6} \noindent
(a) LiAlO$_2$ (100)/Li (001); (b) LiI (001)/Li (001).  
${\cal V}_e$=0.63~V and 1.22~V, respectively.  Pink, red, purple,
and green spheres represent Al, O, Li, and I atoms, respectively.
}
\end{figure}

Alumina can be coated on Cu current collectors, or directly on Li surfaces,
to create aSEIs.\cite{elam,kozen,merrill} Upon cycling, it has been proposed
that alumina reacts to form mixed Al/Li oxides.\cite{elam,kozen,lialox}  We
focus on the mixed oxide candidate LiAlO$_2$.\cite{jacs}  LiAlO$_2$ is not
thermodynamically stable against Li metal; its Al$^{3+}$ can be reduced.
However, we find that its (100) surface is kinetically stable on Li(001)
upon optimization of the interfacial configuration (Fig.~\ref{fig6}a), even
after a short AIMD run (Method).  In contrast, LiI,\cite{lii} like LiH, has a
maximally reduced anion which cannot react with Li metal (Fig.~\ref{fig6}b).

The electronic voltages ${\cal V}_e$ associated with the two interfaces in
Fig.~\ref{fig6}a-b are +0.63~V and 1.29~V, respectively (Eq.~\ref{eq2}).  
These interfaces are above the equilibrium (0.0~V vs.~Li$^+$/Li(s))
conditions.  If a liquid electrolyte were present, Li metal would start to
dissolve as Li$^+$ until $i$-equilibrium is achieved via surface charging.
These values further confirm that the LiF- and Li$_2$O-coated Li metal
near-``open circuit'' conditions are fortuitous exceptions.  Despite this,
the computed ${\cal V}_e$ associated with LiAlO$_2$/Li(s) and LiI/Li(s)
interfaces remain significantly lower than the DFT/PBE predicted bare lithium
``voltage'' of $\sim$1.56~V\cite{gb,pccp} due to the finite contact potentials.
\color{black} Once again, we emphasize that extra effort is needed to keep DFT
model systems at $i$-equilibrium, non-overpotential conditions. \color{black}
Charge-neutral Li$_2$CO$_3$-coated Li surfaces have been predicted to
exhibit ${\cal V}_e$$\sim$0.6~V;\cite{filhol} however, Li$_2$CO$_3$ is in
fact not stable on Li surfaces.\cite{batt,umd} 

The local densities of state (LDOS) plots in the S.I. (Sec.~S3) illustrate the
interfacial band structures.  By construction, the interfaces are not
charged, and as expected, in the insulating SEI/aSEI regions, the valence
band edges (VBE) are flat.  We have also performed spot-checks of the LiI
${\cal V}_e$ with the more accurate HSE06 functional.  The latter increases
${\cal V}_e$ by 0.10~V compared to PBE predicitons (S.I.~Sec.~S3); this
would also be the magnitude of the change in contact potential.

\begin{figure}
\centerline{\hbox{ \epsfxsize=4.00in \epsfbox{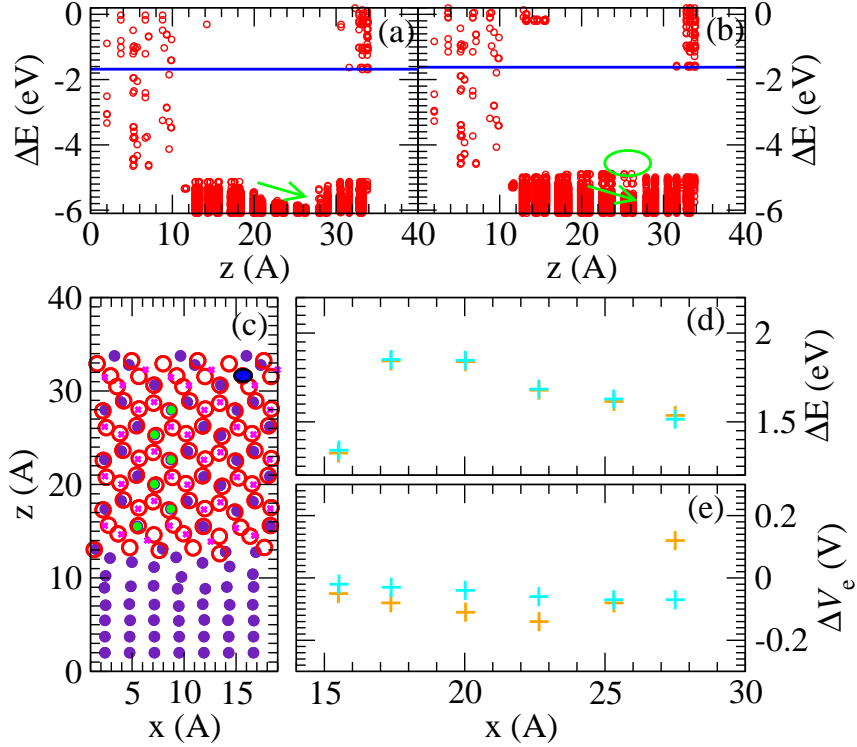} }}
\caption[]
{\label{fig7} \noindent
Li/LiAlO$_2$ with a F substituting a 4-coordinated O~atom near the
LiAlO$_2$/vacuum interface ($z$=31~\AA).  (a)-(b) Local densities of states
for 1$\times$4 supercell, without vacancy and with Li vacancy at 22.5~\AA,
respectively. Green arrows indicate the electric field; the green circle
highlights defect states associated with Li$^+$ vacancy.  Note the partially
occupied surface states at $z$$\sim$33~\AA.  (c) Simulation
cell with F atom position shown as a dark blue sphere.  (d) Cost of
introducing Li vacancy; orange and cyan represent  1$\times$4 and 2$\times$4
supercell results.  (e) Electronic voltage (${\cal V}_e$) associated with the
vacancies in panel (d). 
}
\end{figure}

To illustrate the effect of contact potentials which do not give 
${\cal V}_e$=0.0~V,
we consider Li$^+$ diffusion at the Li/LiAlO$_2$ interface.  First, we attempt
to attain a voltage closer to 0.0~V vs.~Li$^+$/Li(s) than 0.63~V.  This
requires a positively charged surface which induces a negative charge on the
Li metal surfaces, and creates a surface dipole (Eq.~\ref{eq3}) and an electric
field across the LiAlO$_2$ film.  The field inside LiAlO$_2$ is the sole
contribution to the EDL in our model.  Changing the EDL by increasing the
cation surface concentration is also the way voltage is lowered in modeling
pristine electrodes/liquid electrolyte interfaces.\cite{borodin} In
Fig.~\ref{fig7}c, we artificially substitute one or two O~atoms with fluorine
(F) near the LiAlO$_2$/vacuum interface in 1$\times$2 or 1$\times$4 supercells
respectively.  This reduces ${\cal V}_e$ from 0.63~V to 0.31~V, much closer to
equilibrium Li-plating/stripping conditions.  The LDOS (Fig.~\ref{fig7}a)
shows that  the F atom(s) at the surface already creates a ${\cal E}_e$ field,
reflected in the slope of the VBE as $z$ increases (arrow in Fig.~\ref{fig7}a).
It also indicates a small amount of occupied surface states introduced by
the F-atoms.  Adding even more F~atoms does not further lower ${\cal V}_e$
below 0.31~V because $E_{\rm F}$ is pinned by these surface impurity states.
A more realistic model, with organic SEI and liquid electrolytes, appears
necessary to lower ${\cal V}_e$ further.  

Fig.~\ref{fig7}d depicts Li$^+$-vacancy displacement energetics of this model.
Five 1$\times$4 and 2$\times$4 simulation cells with one V$_{\rm Li+}$ each at
different vertical positions are considered (Fig.~\ref{fig7}c).  The Li$^+$
vacancy energy cost ($\Delta E_{\rm Li-vac}$) decreases with distance from the
inner surface (except the vacancy closest to the Li metal surface, which is
an outlier).  The S.I. (Sec.~S7) shows that, without F-substitution such that
${\cal V}_e$=0.63~V, the reverse trend is observed; Li$^+$ vacancy costs
increase with increasing distance from the metal surface.  This trend suggests
that, for this Li/LiAlO$_2$ interface, as the voltage further decreases towards
$i$-equilibrium (${\cal V}_e$=0.0~V), $\Delta E_{\rm Li-vac}$ should decrease
with distance from the Li/LiAlO$_2$ interace even faster than Fig.~\ref{fig7}d.
In other words, Li$^+$ stripping from the Li metal anode through this film
should exhibit a more favorable energy landscape than Li-plating, as long as
only Li$^+$ vacancies in LiAlO$_2$, and not positively charged Li-interstitials,
are involved.  Such an asymmetry has seldom been documented at battery interface
because finite electric fields are difficult to measure and have been largely
neglected in DFT calculations.  However, it may contribute to deviation of
Li plating/stripping exchange current ratio from unity.\cite{butler}
We have not computed Li$^+$ diffusion barriers, but those should also be
affected by an electric field.  A field that hinders Li$^+$ vacancy diffusion
towards the liquid elecrolyte in the SEI should reduce self-discharge.   This
would require a large, positive  contact potential between Li metal and the
surface film, so that at ${\cal V}_e$=0.0~V the Li/film interface is positively
charged.  Thus electric fields may be exploited to accelerate
charging/discharge rate.

Surprisingly, $\Delta E_{\rm Li-vac}$ are well-converged with system size
(Fig.~\ref{fig7})d.  This is unlike the case without F-substitution on the
LiAlO$_2$ surface (S.I.), or with Li$_2$O films Fig.~\ref{fig4}.  Normally this
lack of system size dependence would be the signature of an uncharged point
defect.\cite{corrosion}  However, the LDOS (Fig.~\ref{fig7}b) shows that
a vacancy generates new VBE states, which are signatures of a negatively
charged defect.  We postulate that the partially occupied surface
states are responsible for fast convergence (Fig.~\ref{fig7}a-b).  While the
good system size convergence is computationally convenient, the existence
of surface states is not physical; they should readily react with organic
species if any were present on the surface.   Creating an electric field 
via alternatives to F-substitution will be explored in future work.

\section{Conclusions}

We have examined a Li$_2$O-coated, $\sim$50\% Li-coverage Li/Cu metal
junctions as models of Li metal partially plated on a Cu current collector
in batteries with lithium metal anodes.  Using DFT methods and these
models with explicit interfaces, we show that these Li$|$Cu junctions exhibit
significant spatial inhomogeneity with respect to local electrostatic
potentials.  DOL electrochemical reduction reactions are more favorable
and faster on the Li side of oxide-coated junction.  This suggests that solid
electrolyte interphase (SEI) films form more readily on the Li region
than the Cu region, if the SEI film has the same thickness on both regions.
This is in qualitative agreement with the experimental finding that the final
SEI on is thicker Li-coated regions under lithiun metal anode ``galvanic
corrosion'' conditions.  A 100\% Li-coated Cu surface does not exhibit
such an inhomogeneity or galvanic corrosion signatures.  

We also show that LiH, a recently proposed component of SEI formed from liquid
electrolyte decomposition, can react either chemically or electrochemically
with organic solvent molecules.  The electrochemical degradation process has
a reasonably low reaction barrier consistent with about one~second reaction
time scales.  The chemical route, while exothermic, has a surprisingly high
barrier.  This suggests that LiH may persist in electronically disconnected
regions of the SEI despite its thermodynamic instability.  If LiH is 
continuously formed in the SEI, it may continuously react electrochemically
and deplete Li metal, which becomes relevant to battery self-discharge.  
Finally, we demonstrate that LiAlO$_2$, a model for artificial SEI (aSEI)
on Li metal, should exhibit a cross-film electric field when held at the
plating/stripping potential (0.0~V vs.~Li$^+$/Li(s)).  The field direction
implies that the plating/stripping rates are intrinsically asymmetric.
This feature can potentially be exploited to limit self-discharge batteries.

Our predictions are made possible by previous conceptual advances in DFT
modeling of battery interfaces.  These include voltage definitions
(ionic/equilibrium, ${\cal V}_i$ vs.~electronic/instantaneous, ${\cal V}_e$),
DFT overpotential concepts (${\cal V}_i$$\neq$${\cal V}_e$), accounting for
contact potentials between lithium metal and the inorganic SEI/aSEI, and
including an electric field across the surface film to constrain the system at
a particular ${\cal V}_e$.  ${\cal V}_e$, proportional to the Fermi level, is
shown to be the key determinant of electrochemical reaction energetics when
long-range $e^-$ transfer is involved; the ionic, equilibrium voltage 
${\cal V}_i$, related to Li-insertion energies, is only indirectly relevant.
Calculating ${\cal V}_e$ is also critical for avoiding unintentional
overpotential in DFT calculations that may lead to erroneous comparison
with measurements.

\begin{figure}
\centerline{\hbox{  \epsfxsize=5.50in \epsfbox{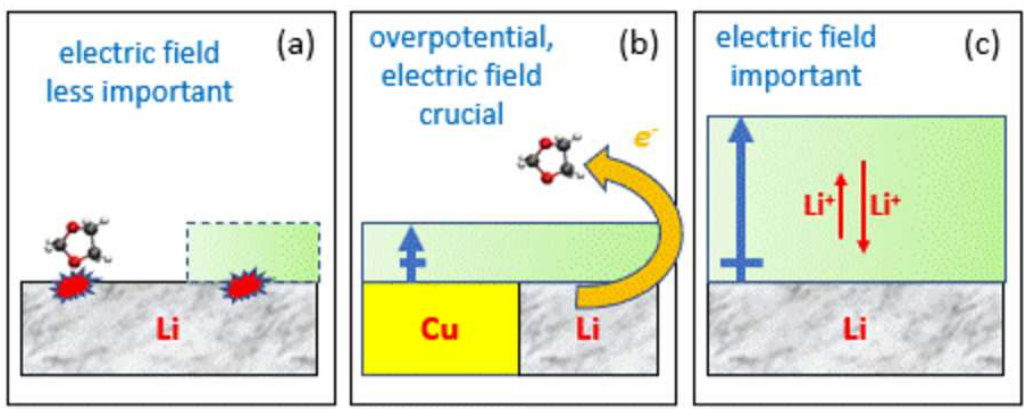} }}
\caption[]
{\label{fig8} \noindent
Schematics of electric field effect at battery anode interfaces.
}
\end{figure}

We find that different artificial SEI or SEI components (Li$_2$O, LiAlO$_2$,
LiI, and LiH) exhibit significantly different contact potentials.  This means
that, at the same applied voltage and film thickness, they would exhibit
electric fields with different magnitudes across the films.  From this and
from our previous work,\cite{pccp,gb,lipon} we postulate that electric fields
have weak influence on chemical reactions between Li metal and the SEI/aSEI
which are directly in contact; are more important for Li$^+$ transport in the
surface film and may therefore affect the stripping/plating rate ratio when
using Butler-Volmer equations;\cite{butler} and are crucial in determining the
in-film voltage drop that governs long range $e^-$ leakage through the
SEI/aSEI, which leads to late-stage SEI growth/evolution (Fig.~\ref{fig8}).
Electric fields inside the SEI are seldom discussed in the battery
literature,\cite{maier} and the contact potential is an even more obscure
topic.  Understanding how such fields (and the accompanying potential
drops/rises) are partitioned among the surface films, the liquid electrolyte,
and the contact potentials is an urgent need that requires further research.

In general, battery interfacial stability, related to cycling as well as
self-discharge, is found to be far more complex than purely
thermodynamic/phase diagram approaches, or standard galvanic corrosion
models for metal corrosion, have postulated.  While our conclusions are
based on somewhat idealized models, we stress that for SEI formation reactions
via long-range electron transfer, the voltage is critical.  Indeed we focus on
voltage-function relations rather than interface structure-function
relations because the atomic details of interfaces have seldom been imaged.
There are fundamental differences between Li metal self-discharge
and corrosion/galvanic corrosion of metals like steel or aluminum, due
to differences in voltage dependence and cation diffusion rate in oxides.
Onset of pitting through breaching of the passivating film at elevated voltages
relative to Li-plating is not required in Li self-discharge, unlike in pitting
corrosion of Al or steel; indeed high potential impedes the cathodic reactions
that lead to Li self-discharge.  The different corrosion mechanisms mean
different mitigating strategies should be applied.  Nevertheless, we propose
that synergistic study of battery interfaces and metal corrosion will yield
significant cross-cutting benefits.

\section*{Acknowledgement}
 
We thank Quinn Campbell for valuable suggestions.
This work is funded by the Laboratory
Directed Research and Development Program at Sandia National Laboratories.  
Sandia National Laboratories is a multi-mission laboratory managed and operated
by National Technology and Engineering Solutions of Sandia, LLC, a wholly owned
subsidiary of Honeywell International, Inc., for the U.S. Department of
Energy’s National Nuclear Security Administration under contract DE-NA0003525.
This paper describes objective technical results and analysis.  Any subjective
views or opinions that might be expressed in the document do not necessarily
represent the views of the U.S. Department of Energy or the United States
Government.

\section*{Supporting Information}
The Supporting Information is available free of charge at 
https://pubs.acs.org/.

Experimental evidence for Li$_2$O; experimental galvanic corrosion data;
predicted local densities of state; DOL decomposition in solution and
on Li surface; comparison with other battery interface DFT work; discussion
of voltage definitions; Li$^+$ vacancy diffusion without surface F-dopant;
simulation cell details; DOL decomposition on Li$_2$O surface; additional
predictions on an``oxide~3'' model; and surface/interfacial contributions
to voltage


\begin{references}
\bibitem{note0}
Our solid-solid interface focus is not meant to suggest the liquid
electrolytes do not play a critical role in battery stabiilty; high
salt concentration electrolytes,\cite{highconc,highconc2,wise} anions
effects,\cite{baskin} and electrolyte additives which yield different
aSEI components, like LiI,\cite{lii} are examples of liquid electrolyte
effects.
\end{references}
\end{document}